\documentclass{kapedbk}
\usepackage[dvips]{graphicx}

\newcommand{\url}[1]{{\texttt{#1}}}
\newcommand{\Prob}{\mathrm{Pr}}
\newcommand{\Dist}{\mathrm{P}}
\newcommand{\AltDist}{\mathrm{Q}}
\newcommand{\Expec}[1]{\mathbf{E}\left[#1\right]}
\newcommand{\argmin}[2]{\mathrm{arg}\!\min_{#1}{#2}}
\newcommand{\bold}{\bf}
\newcommand{\TimeAvg}[1]{\ensuremath{\left \langle #1 \right \rangle}}

\upperandlowercase
\normallatexbib 

\begin{document}
\articletitle[Overview of Methods and Techniques]{Methods and Techniques of Complex Systems Science: An Overview}
\author{Cosma Rohilla Shalizi}
\affil{Center for the Study of Complex Systems, University of Michigan, Ann Arbor, MI 48109 USA}
\email{cshalizi@umich.edu}

\begin{abstract}
In this chapter, I review the main methods and techniques of complex systems
science.  As a first step, I distinguish among the broad patterns which recur
across complex systems, the topics complex systems science commonly studies,
the tools employed, and the foundational science of complex systems.  The focus
of this chapter is overwhelmingly on the third heading, that of tools.  These
in turn divide, roughly, into tools for analyzing data, tools for constructing
and evaluating models, and tools for measuring complexity.  I discuss the
principles of statistical learning and model selection; time series analysis;
cellular automata; agent-based models; the evaluation of complex-systems
models; information theory; and ways of measuring complexity.  Throughout, I
give only rough outlines of techniques, so that readers, confronted with new
problems, will have a sense of which ones might be suitable, and which ones
definitely are not.
\end{abstract}

\section{Introduction}

\begin{figure}
\begin{center}
\resizebox{2.0in}{!}{\includegraphics{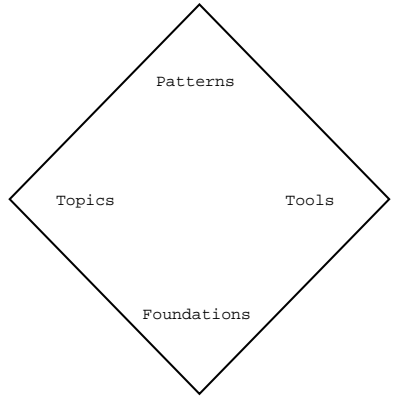}}
\end{center}
\caption{The quadrangle of complex systems.  See text.}
\label{figure:quadrangle}
\end{figure}

A complex system, roughly speaking, is one with many parts, whose behaviors are
both highly variable and strongly dependent on the behavior of the other parts.
Clearly, this includes a large fraction of the universe!  Nonetheless, it is
not vacuously all-embracing: it excludes both systems whose parts just cannot
do very much, and those whose parts are really independent of each other.
``Complex systems science'' is the field whose ambition is to understand
complex systems.  Of course, this is a broad endeavor, overlapping with many
even larger, better-established scientific fields.  Having been asked by the
editors to describe its methods and techniques, I begin by explaining what I
feel does {\em not} fall within my charge, as indicated by Figure \ref{figure:quadrangle}.

\begin{figure}
\begin{center}
\resizebox{2in}{!}{\includegraphics{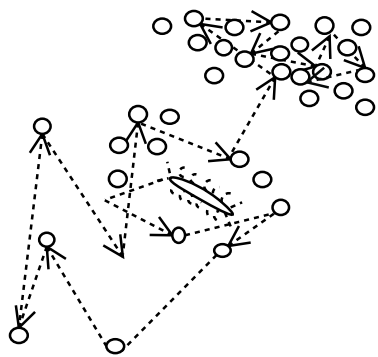}}
\end{center}
\caption{Bacterial chemotaxis.  Should the bacterium (center) exploit the
  currently-available patch of food, or explore, in hopes of finding richer
  patches elsewhere (e.g. at right)?  Many species solve this problem by
  performing a random walk (jagged line), tumbling randomly every so often.
  The frequency of tumbling increases when the concentration of nutrients is
  high, making the bacterium take long steps in resource-poor regions, and
  persist in resource-rich ones
  \cite{Strong-et-al-optimal-chemotaxis,Alon-et-al-robustness-of-chemotaxis,%
    Yi-et-al-perfect-adaptation-in-chemotaxis}.}
\label{fig:chemotaxis}
\end{figure}

At the top of Figure \ref{figure:quadrangle} I have put ``patterns''.  By this
I mean more or less what people in software engineering do
\cite{Design-Patterns-gang-of-four}: a pattern is a recurring theme in the
analysis of many different systems, a cross-systemic regularity.  For instance:
bacterial chemotaxis can be thought of as a way of resolving the tension
between the exploitation of known resources, and costly exploration for new,
potentially more valuable, resources (Figure \ref{fig:chemotaxis}).  This same
tension is present in a vast range of adaptive systems.  Whether the
exploration-exploitation trade-off arises among artificial agents, human
decision-makers or colonial organisms, many of the issues are the same as in
chemotaxis, and solutions and methods of investigation that apply in one case
can profitably be tried in another
\cite{Anderson-random-walk-learning,Mueller-et-al-optimization-via-chemotaxis}.
The pattern ``trade-off between exploitation and exploration'' thus serves to
orient us to broad features of novel situations.  There are many other such
patterns in complex systems science: ``stability through hierarchically
structured interactions'' \cite{Simon-architecture}, ``positive feedback
leading to highly skewed outcomes'' \cite{Simon-skew}, ``local inhibition and
long-rate activation create spatial patterns'' \cite{Turing-morphogenesis}, and
so forth.

At the bottom of the quadrangle is ``foundations'', meaning attempts to build a
basic, mathematical science concerned with such topics as the measurement of
complexity \cite{Badii-Politi}, the nature of organization
\cite{Fontana-Buss-organization}, the relationship between physical processes
and information and computation \cite{complexity-entropy-physics-of-info} and
the origins of complexity in nature and its increase (or decrease) over time.
There is dispute whether such a science is possible, if so whether it would be
profitable.  I think it is both possible and useful, but most of what has been
done in this area is very far from being applicable to {\em biomedical}
research.  Accordingly, I shall pass it over, with the exception of a brief
discussion of some work on measuring complexity and organization which is
especially closely tied to data analysis.

``Topics'' go in the left-hand corner.  Here are what one might call the
``canonical complex systems'', the particular systems, natural, artificial and
fictional, which complex systems science has traditionally and habitually
sought to understand.  Here we find networks (Wuchty, Ravasz and Barab{\'a}si,
this volume), turbulence \cite{Frisch-turbulence}, physio-chemical pattern
formation and biological morphogenesis \cite{Cross-Hohenberg,Ball-tapestry},
genetic algorithms \cite{Holland-adaptation,MM-on-GAs}, evolutionary dynamics
\cite{Gintis-game-theory-evolving,Hofbauer-Sigmund}, spin glasses
\cite{Fischer-and-Hertz-spin-glasses,Stein-spin-glasses-still-complex},
neuronal networks (see Part III, 4, this book), the immune system (see Part
III, 5, this book), social insects, ant-like robotic systems, the evolution of
cooperation, evolutionary economics, etc.\footnote{Several books pretend to
  give a unified presentation of the topics.  To date, the only one worth
  reading is \cite{Boccara-complex-systems}, which however omits all models of
  adaptive systems.}  These topics all fall within our initial definition of
``complexity'', though whether they are studied together because of deep
connections, or because of historical accidents and tradition, is a difficult
question.  In any event, this chapter will {\em not} describe the facts and
particular models relevant to these topics.

Instead, this chapter is about the right-hand corner, ``tools''.  Some are
procedures for analyzing data, some are for constructing and evaluating models,
and some are for measuring the complexity of data or models.  In this chapter I
will restrict myself to methods which are generally accepted as valid (if not
always widely applied), {\em and} seem promising for biomedical research.
These still demand a book, if not an encyclopedia, rather than a mere chapter!
Accordingly, I will merely try to convey the essentials of the methods, with
pointers to references for details.  The goal is for you to have a sense of
which methods would be good things to try on your problem, rather than to tell
you everything you need to know to implement them.

\subsection{Outline of This Chapter}

As mentioned above, the techniques of complex systems science can, for our
purposes, be divided into three parts: those for analyzing data (perhaps
without reference to a particular model), those for building and understanding
models (often without data), and those for measuring complexity as such.  This
chapter will examine them in that order.

The first part, on {\bold data}, opens with the general ideas of {\bold
  statistical learning and data mining} (\S \ref{sec:data-mining}), namely
developments in statistics and machine learning theory that extend statistical
methods beyond their traditional domain of low-dimensional, independent data.
We then turn to {\bold time series analysis} (\S \ref{sec:time-series}), where
there are two important streams of work, inspired by statistics and nonlinear
dynamics.

The second part, on {\bold modeling}, considers the most important and
distinctive classes of models in complex systems.  On the vital area of {\bold
  nonlinear dynamics}, let the reader consult Socoloar (this volume).  {\bold
  Cellular automata} (\S \ref{sec:cas}) allow us to represent spatial dynamics
in a way which is particularly suited to capturing strong local interactions,
spatial heterogeneity, and large-scale aggregate patterns.  Complementary to
cellular automata are {\bold agent-based models} (\S \ref{sec:abms}), perhaps
the most distinctive and most famous kind of model in complex systems science.
A general section (\ref{sec:evaluation}) on {\bold evaluating complex models},
including analytical methods, various sorts of simulation, and testing, closes
this part of the chapter.

The third part of the chapter considers ways of measuring complexity.  As a
necessary preliminary, \S \ref{sec:info-theory} introduces the concepts of
{\bold information theory}, with some remarks on its application to biological
systems.  Then \S \ref{sec:complexity} treats {\bold complexity measures},
describing the main kinds of complexity measure, their relationships, and their
applicability to empirical questions.

The chapter ends with a guide to further reading, organized by section.  These
emphasize readable and thorough introductions and surveys over more advanced or
historically important contributions.

\section{Statistical Learning and Data-Mining}
\label{sec:data-mining}

Complex systems, we said, are those with many strongly interdependent parts.
Thanks to comparatively recent developments in statistics and machine learning,
it is now possible to infer reliable, predictive models from data, even when
the data concern thousands of strongly dependent variables.  Such {\bold data
  mining} is now a routine part of many industries, and is increasingly
important in research.  While not, of course, a substitute for devising valid
theoretical models, data mining {\em can} tell us what kinds of patterns are in
the data, and so guide our model-building.

\subsection{Prediction and Model Selection}

The basic goal of any kind of data mining is prediction: some variables, let us
call them $X$, are our inputs.  The output is another variable or variables
$Y$.  We wish to use $X$ to predict $Y$, or, more exactly, we wish to build a
machine which will do the prediction for us: we will put in $X$ at one end, and
get a prediction for $Y$ out at the other.\footnote{Not all data mining is
  strictly for predictive models.  One can also mine for purely descriptive
  models, which try to, say, cluster the data points so that more similar ones
  are closer together, or just assign an over-all likelihood score.  These,
  too, can be regarded as minimizing a cost function (e.g., the dis-similarity
  within clusters plus the similarity across clusters).  The important point is
  that good descriptions, in this sense, are implicitly predictive, either
  about other aspects of the data or about further data from the same source.}

``Prediction'' here covers a lot of ground.  If $Y$ are simply other variables
like $X$, we sometimes call the problem {\bold regression}.  If they are $X$ at
another time, we have {\bold forecasting}, or prediction in a strict sense of
the word.  If $Y$ indicates membership in some set of discrete categories, we
have {\bold classification}.  Similarly, our predictions for $Y$ can take the
form of distinct, particular values ({\bold point predictions}), of ranges or
intervals we believe $Y$ will fall into, or of entire probability distributions
for $Y$, i.e., guesses as to the conditional distribution $\Prob(Y|X)$.  One
can get a point prediction from a distribution by finding its mean or mode, so
distribution predictions are in a sense more complete, but they are also more
computationally expensive to make, and harder to make successfully.

Whatever kind of prediction problem we are attempting, and with whatever kind
of guesses we want our machine to make, we must be able to say whether or not
they are good guesses; in fact we must be able to say just how much bad guesses
cost us.  That is, we need a {\bold loss function} for predictions\footnote{A
  subtle issue can arise here, in that not all errors need be equally bad for
  us.  In scientific applications, we normally aim at accuracy as such, and so
  all errors {\em are} equally bad.  But in other applications, we might care
  very much about otherwise small inaccuracies in some circumstances, and shrug
  off large inaccuracies in others.  A well-designed loss function will
  represent these desires.  This does not change the basic principles of
  learning, but it can matter a great deal for the final machine
  \cite{Spyros-decisionmetrics}.}.  We suppose that our machine has a number of
knobs and dials we can adjust, and we refer to these parameters, collectively,
as $\theta$.  The predictions we make, with inputs $X$ and parameters $\theta$,
are $f(X,\theta)$, and the loss from the error in these predictions, when the
actual outputs are $Y$, is $L(Y,f(X,\theta))$.  Given {\em particular} values
$y$ and $x$, we have the empirical loss $L(y,f(x,\theta))$, or
$\hat{L}(\theta)$ for short\footnote{Here and throughout, I try to follow the
  standard notation of probability theory, so capital letters ($X, Y$, etc.)
  denote random variables, and lower-case ones particular values or
  realizations --- so $X$ = the role of a die, whereas $x = 5$ (say).}.

Now, a natural impulse at this point is to twist the knobs to make the loss
small: that is, to select the $\theta$ which minimizes $\hat{L}(\theta)$; let's
write this $\hat{\theta} = \argmin{\theta}{\hat{L}(\theta)}$.  This procedure
is sometimes called {\bold empirical risk minimization}, or ERM.  (Of course,
doing that minimization can itself be a tricky nonlinear problem, but I will
not cover optimization methods here.)  The problem with ERM is that the
$\hat{\theta}$ we get from {\em this} data will almost surely not be the same
as the one we'd get from the {\em next} set of data.  What we really care
about, if we think it through, is not the error on any particular set of data,
but the error we can {\em expect} on new data, $\Expec{L(\theta)}$.  The
former, $\hat{L}(\theta)$, is called the {\bold training} or {\bold in-sample}
or {\bold empirical} error; the latter, $\Expec{L(\theta)}$, the {\bold
  generalization} or {\bold out-of-sample} or {\bold true} error.  The
difference between in-sample and out-of-sample errors is due to sampling noise,
the fact that our data are not {\em perfectly} representative of the system
we're studying.  There will be quirks in our data which are just due to chance,
but if we minimize $\hat{L}$ blindly, if we try to reproduce every feature of
the data, we will be making a machine which reproduces the random quirks, which
do not generalize, along with the predictive features.  Think of the empirical
error $\hat{L}(\theta)$ as the generalization error, $\Expec{L(\theta)}$, plus
a sampling fluctuation, $\epsilon$.  If we look at machines with low empirical
errors, we will pick out ones with low true errors, which is good, but we will
also pick out ones with large negative sampling fluctuations, which is not
good.  Even if the sampling noise $\epsilon$ is very small, $\hat{\theta}$ can
be very different from $\theta_{min}$.  We have what optimization theory calls
an {\bold ill-posed problem} \cite{Vapnik-nature}.

Having a higher-than-optimal generalization error because we paid too much
attention to our data is called {\bold over-fitting}.  Just as we are often
better off if we tactfully ignore our friends' and neighbors' little faults, we
want to ignore the unrepresentative blemishes of our sample.  Much of the
theory of data mining is about avoiding over-fitting.  Three of the commonest
forms of tact it has developed are, in order of sophistication, {\bold
  cross-validation}, {\bold regularization} (or {\bold penalties}) and {\bold
  capacity control}.

\subsubsection{Validation}

We would never over-fit if we {\em knew} how well our machine's predictions
would generalize to new data.  Since our data is never perfectly
representative, we always have to estimate the generalization performance.  The
empirical error provides one estimate, but it's biased towards saying that the
machine will do well (since we built it to do well on that data).  If we had a
second, independent set of data, we could evaluate our machine's predictions on
it, and that would give us an unbiased estimate of its generalization.  One way
to do this is to take our original data and divide it, at random, into two
parts, the {\bold training set} and the {\bold test set} or {\bold validation}
set.  We then use the training set to fit the machine, and evaluate its
performance on the test set.  (This is an instance of {\bold resampling} our
data, which is a useful trick in many contexts.)  Because we've made sure the
test set is independent of the training set, we get an unbiased estimate of the
out-of-sample performance.

In {\bold cross-validation}, we divide our data into random training and test
sets many different ways, fit a different machine for each training set, and
compare their performances on their test sets, taking the one with the best
test-set performance.  This re-introduces some bias --- it could happen by
chance that one test set reproduces the sampling quirks of its training set,
favoring the model fit to the latter.  But cross-validation generally {\em
  reduces} over-fitting, compared to simply minimizing the empirical error; it
makes more {\em efficient} use of the data, though it cannot get rid of
sampling noise altogether.

\subsubsection{Regularization or Penalization}

I said that the problem of minimizing the error is {\bold ill-posed}, meaning
that small changes in the errors can lead to big changes in the optimal
parameters.  A standard approach to ill-posed problems in optimization theory
is called {\bold regularization}.  Rather than trying to minimize
$\hat{L}(\theta)$ alone, we minimize
\begin{equation}
\hat{L}(\theta) + \lambda d(\theta) ~,
\label{eqn:regularization}
\end{equation}
where $d(\theta)$ is a {\bold regularizing} or {\bold penalty} function.
Remember that $\hat{L}(\theta) = \Expec{L(\theta)} + \epsilon$, where
$\epsilon$ is the sampling noise.  If the penalty term is well-designed, then
the $\theta$ which minimizes
\begin{equation}
\Expec{L(\theta)} + \epsilon + \lambda d(\theta)
\label{eqn:regularization-incl-noise}
\end{equation}
will be close to the $\theta$ which minimizes $\Expec{L(\theta)}$ --- it will
cancel out the effects of favorable fluctuations.  As we acquire more and more
data, $\epsilon \rightarrow 0$, so $\lambda$, too, goes to zero at an
appropriate pace, the penalized solution will converge on the machine with the
best possible generalization error.

How then should we design penalty functions?  The more knobs and dials there
are on our machine, the more opportunities we have to get into mischief by
matching chance quirks in the data.  If one machine with fifty knobs, and
another fits the data just as well but has only a single knob, we should (the
story goes) chose the latter --- because it's {\em less} flexible, the fact
that it does well is a good indication that it will still do well in the
future.  There are thus many regularization methods which add a penalty
proportional to the number of knobs, or, more formally, the number of
parameters.  These include the Akaike information criterion or AIC
\cite{Akaike-AIC} and the Bayesian information criterion or BIC
\cite{Akaike-BIC,Schwarz-BIC}.  Other methods penalized the ``roughness'' of a
model, i.e., some measure of how much the prediction shifts with a small change
in either the input or the parameters \cite[ch.\ 10]{van-de-Geer-empirical}.  A
smooth function is less flexible, and so has less ability to match meaningless
wiggles in the data.  Another popular penalty method, the {\bold minimum
  description length} principle of Rissanen, will be dealt with in \S
\ref{sec:stat-compl} below.

\begin{figure}
\begin{center}
\resizebox{3in}{!}{\includegraphics{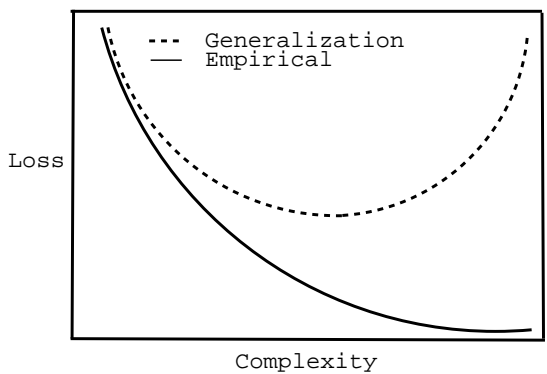}}
\end{center}
\caption{Empirical loss and generalization loss as a function of model
  complexity.}
\label{fig:overfit}
\end{figure}

Usually, regularization methods are justified by the idea that models can be
more or less complex, and more complex ones are more liable to over-fit, all
else being equal, so penalty terms should reflect complexity (Figure
\ref{fig:overfit}).  There's something to this idea, but the usual way of
putting it does not really work; see \S \ref{sec:razor} below.

\subsubsection{Capacity Control}
\label{sec:PAC-and-SRM}

Empirical risk minimization, we said, is apt to over-fit because we do not know
the generalization errors, just the empirical errors.  This would not be such a
problem if we could {\em guarantee} that the in-sample performance was close to
the out-of-sample performance.  Even if the exact machine we got this way
was not particularly close to the optimal machine, we'd then be guaranteed that
our {\em predictions} were nearly optimal.  We do not even need to guarantee
that {\em all} the empirical errors are close to their true values, just that
the {\em smallest} empirical error is close to the smallest generalization
error.

Recall that $\hat{L}(\theta) = \Expec{L(\theta)} + \epsilon$.  It is natural to
assume that as our sample size $N$ becomes larger, our sampling error
$\epsilon$ will approach zero.  (We will return to this assumption below.)
Suppose we could find a function $\eta(N)$ to bound our sampling error, such
that $|\epsilon| \leq \eta(N)$.  Then we could guarantee that our choice of
model was {\bold approximately correct}; if we wanted to be sure that our
prediction errors were within $\epsilon$ of the best possible, we would merely
need to have $N(\epsilon) = \eta^{-1}(\epsilon)$ data-points.

It should not be surprising to learn that we cannot, generally, make
approximately correct guarantees.  As the eminent forensic statistician C. Chan
remarked, ``Improbable events permit themselves the luxury of occurring''
\cite{Behind-that-curtain}, and one of these indulgences could make the
discrepancy between $\hat{L}(\theta)$ and $\Expec{L(\theta)}$ very large.  But
if something like the law of large numbers holds, or the ergodic theorem (\S
\ref{sec:time-series-properties}), then for every choice of $\theta$,
\begin{eqnarray}
\Prob\left(\left|\hat{L}(\theta) - \Expec{L(\theta)}\right| > \epsilon\right) &
\rightarrow & 0 ~,
\end{eqnarray}
for every positive $\epsilon$.\footnote{This is called the {\bold convergence
    in probability} of $\hat{L}(\theta)$ to its mean value.  For a practical
  introduction to such convergence properties, the necessary and sufficient
  conditions for them to obtain, and some thoughts about what one can do,
  statistically, when they do not, see \cite{Gray-ergodic-properties}.} We
should be able to find some function $\delta$ such that
\begin{eqnarray}
\Prob\left(\left|\hat{L}(\theta) - \Expec{L(\theta)}\right| > \epsilon\right) &
\leq & \delta(N,\epsilon,\theta) ~,
\end{eqnarray}
with $\lim_{N}{\delta(N,\epsilon,\theta)} = 0$.  Then, for any particular
$\theta$, we could give {\bold probably approximately correct}
\cite{Valiant-learnable} guarantees, and say that, e.g., to have a 95\%
confidence that the true error is within 0.001 of the empirical error requires
at least 144,000 samples (or whatever the precise numbers may be).  If we can
give probably approximately correct (PAC) guarantees on the performance of one
machine, we can give them for any {\em finite} collection of machines.  But if
we have infinitely many possible machines, might not there always be {\em some}
of them which are misbehaving?  Can we still give PAC guarantees when $\theta$
is continuous?

The answer to this question depends on how flexible the set of machines is ---
its {\bold capacity}.  We need to know how easy it is to find a $\theta$ such
that $f(X,\theta)$ will accommodate itself to any $Y$.  This is measured by a
quantity called the Vapnik-Chervonenkis (VC) dimension
\cite{Vapnik-nature}\footnote{The precise definition of the VC dimension is
  somewhat involved, and omitted here for brevity's sake.  See
  \cite{Kearns-Vazirani,Cristianini-Shawe-Taylor} for clear discussions.}.  If
the VC dimension $d$ of a class of machines is finite, one can make a PAC
guarantee which applies to {\em all} machines in the class simultaneously:
\begin{eqnarray}
\Prob\left(\max_{\theta}{\left|\hat{L}(\theta) - \Expec{L(\theta)}\right|} \geq \eta(N,d,\delta)\right) & \leq & \delta
\label{VC-PAC}
\end{eqnarray}
where the function $\eta(N,d,\delta)$ expresses the rate of convergence.  It
depends on the particular kind of loss function involved.  For example, for
binary classification, if the loss function is the fraction of inputs
mis-classified,
\begin{eqnarray}
\label{classification-bound}
\eta(N,d,\delta) & = & \frac{1}{\sqrt{N}}\sqrt{\left(d(1+\ln{\frac{2N}{d}}) +
  \ln{\frac{4}{\delta}}\right)}
\end{eqnarray}
Notice that $\theta$ is not an argument to $\eta$, and does not appear in
(\ref{classification-bound}).  The rate of convergence is the same across all
machines; this kind of result is thus called a {\bold uniform law of large
  numbers}.  The really remarkable thing about (\ref{VC-PAC}) is that it holds
no matter what the sampling distribution is, so long as samples are
independent; it is a {\bold distribution-free} result.

The VC bounds lead to a very nice learning scheme: simply apply empirical risk
minimization, for a fixed class of machines, and then give a PAC guarantee that
the one picked is, with high reliability, very close to the actual optimal
machine.  The VC bounds also lead an appealing penalization scheme, where the
penalty is equal to our bound on the over-fitting, $\eta$.  Specifically, we
set the $\lambda d(\theta)$ term in (\ref{eqn:regularization}) equal to the
$\eta$ in (\ref{VC-PAC}), ensuring, with high probability, that the $\epsilon$
and $\lambda d(\theta)$ terms in (\ref{eqn:regularization-incl-noise}) cancel
each other.  This is {\bold structural risk minimization} (SRM).

It's important to realize that the VC dimension is not the same as the number
of parameters.  For some classes of functions, it is much {\em lower} than the
number of parameters, and for others it's much {\em higher}.  (There are
examples of one-parameter classes of functions with infinite VC dimension.)
Determining the VC dimension often involves subtle combinatorial arguments, but
many results are now available in the literature, and more are appearing all
the time.  There are even schemes for experimentally estimating the VC
dimension \cite{Shao-et-al-measuring-VC-dim}.

Two caveats are in order.  First, because the VC bounds are distribution-free,
they are really about the rate of convergence under the worst possible
distribution, the one a malicious Adversary out to foil our data mining would
choose.  This means that in practice, convergence is often much faster than
(\ref{VC-PAC}) would indicate.  Second, the usual proofs of the VC bounds all
assume independent, identically-distributed samples, though the relationship
between $X$ and $Y$ can involve arbitrarily complicated
dependencies\footnote{For instance, one can apply the independent-sample theory
  to learning feedback control systems \cite{Vidyasagar-learning}.}. Recently,
there has been much progress in proving uniform laws of large numbers for
dependent sequences of samples, and structural risk minimization has been
extended to what are called ``mixing'' processes
\cite{Meir-nonparametric-time-series}, in effect including an extra term in the
$eta$ function appearing in (\ref{VC-PAC}) which discounts the number of
observations by their degree of mutual dependence.

\subsection{Choice of Architecture}
\label{sec:architectures}

The basic idea of data mining is to fit a model to data with minimal
assumptions about what the correct model should be, or how the variables in the
data are related.  (This differs from such classical statistical questions as
testing {\em specific} hypotheses about specific models, such as the presence
of interactions between certain variables.)  This is facilitated by the
development of extremely flexible classes of models, which are sometimes,
misleadingly, called {\bold non-parametric}; a better name would be {\bold
  megaparametric}.  The idea behind megaparametric models is that they should
be capable of approximating any function, at least any well-behaved function,
to any desired accuracy, given enough capacity.

The polynomials are a familiar example of a class of functions which can
perform such universal approximation.  Given any smooth function $f$, we can
represent it by taking the Taylor series around our favorite point $x_0$.
Truncating that series gives an approximation to $f$:
\begin{eqnarray}
f(x) & = & f(x_0) + \sum_{k=1}^{\infty}{\frac{{(x-x_0)}^k}{k!}{\left.\frac{d^k f}{dx^k}\right|}_{x_0}}\\
& \approx & f(x_0) + \sum_{k=1}^{n}{\frac{{(x-x_0)}^k}{k!}{\left.\frac{d^k f}{dx^k}\right|}_{x_0}}\\
& = & \sum_{k=0}^{n}{a_k \frac{{(x-x_0)}^k}{k!}}
\end{eqnarray}
In fact, if $f$ is an $n^{\mathrm th}$ order polynomial, the truncated series
is exact, not an approximation.

To see why this is {\em not} a reason to use only polynomial models, think
about what would happen if $f(x) = \sin{x}$.  We would need an {\em infinite}
order polynomial to completely represent $f$, and the generalization properties
of finite-order approximations would generally be lousy: for one thing, $f$ is
bounded between -1 and 1 everywhere, but any finite-order polynomial will start
to zoom off to $\infty$ or $-\infty$ outside some range.  Of course, this $f$
would be really easy to approximate as a superposition of sines and cosines,
which is another class of functions which is capable of universal approximation
(better known, perhaps, as Fourier analysis).  What one wants, naturally, is to
chose a model class which gives a good approximation of the function at hand,
{\em at low order}.  We want low order functions, both because computational
demands rise with model order, {\em and} because higher order models are more
prone to over-fitting (VC dimension generally rises with model order).

To adequately describe all of the {\em common} model classes, or {\bold model
  architectures}, used in the data mining literature would require another
chapter.  (\cite{tEoSL} and \cite{Ripley-pattern-recognition} are good for
this.)  Instead, I will merely name a few.
\begin{itemize}
\item {\bold Splines} are piecewise polynomials, good for regression on bounded
  domains; there is a very elegant theory for their estimation
  \cite{Wahba-spline-models}.
\item {\bold Neural networks} or {\bold multilayer perceptrons} have a devoted
  following, both for regression and classification
  \cite{Ripley-pattern-recognition}.  The application of VC theory to them is
  quite well-advanced
  \cite{Anthony-Bartlett-neural-network-learning,Zapranis-Refenes}, but there
  are many other approaches, including ones based on statistical mechanics
  \cite{Engel-and-Van-den-Broeck}.  It is notoriously hard to understand {\em
    why} they make the predictions they do.
\item {\bold Classification and regression trees} (CART), introduced in the
  book of that name \cite{CART-book}, recursively sub-divide the input space,
  rather like the game of ``twenty questions'' (``Is the temperature above 20
  centigrade?  If so, is the glucose concentration above one millimole?'',
  etc.); each question is a branch of the tree.  All the cases at the end of
  one branch of the tree are treated equivalently.  The resulting decision
  trees are easy to understand, and often similar to human decision heuristics
  \cite{Gigerenzer-Todd-heuristics}.
\item {\bold Kernel machines}
  \cite{Vapnik-nature,Herbrich-learning-kernel-classifiers} apply nonlinear
  transformations to the input, mapping it to a much higher dimensional
  ``feature space'', where they apply linear prediction methods.  The trick
  works because the VC dimension of linear methods is low, even in
  high-dimensional spaces.  Kernel methods come in many flavors, of which the
  most popular, currently, are {\bold support vector machines}
  \cite{Cristianini-Shawe-Taylor}.
\end{itemize}

\subsubsection{Predictive versus Causal Models}

Predictive and descriptive models both are not necessarily causal.  PAC-type
results give us reliable prediction, {\em assuming} future data will come from
the {\em same} distribution as the past.  In a causal model, however, we want
to know how {\em changes} will propagate through the system.  One difficulty is
that these relationships are one-way, whereas prediction is two-way (one can
predict genetic variants from metabolic rates, but one cannot change genes by
changing metabolism).  The other is that it is hard (if not impossible) to tell
if the predictive relationships we have found are {\bold confounded} by the
influence of other variables and other relationships we have neglected.
Despite these difficulties, the subject of {\bold causal inference} from data
is currently a very active area of research, and many methods have been
proposed, generally under assumptions about the absence of feedback
\cite{Pearl-causality,Shafer-causal-conjecture,Spirtes-Glymour-Scheines}.  When
we have a causal or generative model, we can use very well-established
techniques to infer the values of the hidden or latent variables in the model
from the values of their observed effects
\cite{Pearl-causality,Helmholtz-machine}.

\subsection{Occam's Razor and Complexity in Prediction}
\label{sec:razor}

Often, regularization methods are thought to be penalizing the {\em complexity}
of the model, and so implementing some version of Occam's Razor.  Just as Occam
said ``entities are not to be multiplied beyond necessity''\footnote{Actually,
  the principle goes back to Aristotle at least, and while Occam used it often,
  he never used exactly those words \cite[translator's introduction]{Occam}.},
we say ``parameters should not be multiplied beyond necessity'', or, ``the
model should be no rougher than necessary''.  This takes complexity to be a
property of an {\em individual} model, and the hope is that a simple model
which can predict the training data will also be able to predict new data.
Now, under many circumstances, one can prove that, as the size of the sample
approaches infinity, regularization will converge on the correct model, the one
with the best generalization performance \cite{van-de-Geer-empirical}.  But one
can often prove exactly the same thing about ERM without any regularization or
penalization at all; this is what the VC bounds (\ref{VC-PAC}) accomplish.
While regularization methods often do well in practice, so, too, does straight
ERM.  If we compare the performance of regularization methods to straight
empirical error minimization on artificial examples, where we can calculate the
generalization performance exactly, regularization conveys {\em no clear
  advantage at all} \cite{Domingos-on-Occams-Razor}.

Contrast this with what happens in structural risk minimization.  There our
complexity penalty depends solely on the VC dimension of the {\em class} of
models we're using.  A simple, inflexible model which we find only because
we're looking at a complex, flexible class is penalized just as much as the
most wiggly member of that class.  Experimentally, SRM {\em does} work better
than simple ERM, or than traditional penalization methods.

A simple example may help illuminate why this is so.  Suppose we're interested
in binary classification, and we find a machine $\theta$ which correctly
classifies a million independent data points.  If the real error rate (=
generalization error) for $\theta$ was one in a hundred thousand, the chance
that it would correctly classify a million data points would be
${(0.99999)}^{{10}^{6}} \approx 4.5 \cdot {10}^{-5}$.  If $\theta$ was the very
first parameter setting we checked, we could be quite confident that its true
error rate was much less than ${10}^{-5}$, no matter how complicated the
function $f(X,\theta)$ looked.  But if we've looked at ten million parameter
settings before finding $\theta$, then the odds are quite good that, among the
machines with an error rate of ${10}^{-5}$, we'd find several which correctly
classify all the points in the training set, so the fact that $\theta$ does is
not good evidence that it's the best machine\footnote{This is very close to the
  notion of the power of a statistical hypothesis test \cite{Lehmann-testing},
  and almost exactly the same as the severity of such a test
  \cite{Mayo-error}.}.  What matters is not how much algebra is involved in
making the predictions once we've chosen $\theta$, but how many alternatives to
$\theta$ we've tried out and rejected.  The VC dimension lets us apply this
kind of reasoning rigorously and without needing to know the details of the
process by which we generate and evaluate models.

The upshot is that the kind of complexity which matters for learning, and so
for Occam's Razor, is the complexity of {\em classes of models}, not of
individual models nor of the system being modeled.  It is important to keep
this point in mind when we try to measure the complexity of systems (\S
\ref{sec:complexity}).

\subsection{Relation of Complex Systems Science to Statistics}

Complex systems scientists often regard the field of statistics as irrelevant
to understanding such systems.  This is understandable, since the exposure most
scientists have to statistics (e.g., the ``research methods'' courses
traditional in the life and social sciences) typically deal with systems with
only a few variables and with explicit assumptions of independence, or only
very weak dependence.  The kind of modern methods we have just seen, amenable
to large systems and strong dependence, are rarely taught in such courses, or
even mentioned.  Considering the shaky grasp many students have on even the
basic principles of statistical inference, this is perhaps wise.  Still, it
leads to even quite eminent researchers in complexity making disparaging
remarks about statistics (e.g., ``statistical hypothesis testing, that
substitute for thought''), while actually re-inventing tools and concepts which
have long been familiar to statisticians.

For their part, many statisticians tend to overlook the very {\em existence} of
complex systems science as a separate discipline.  One may hope that the
increasing interest from both fields on topics such as bioinformatics and
networks will lead to greater mutual appreciation.

\section{Time Series Analysis}
\label{sec:time-series}

There are two main schools of time series analysis.  The older one, which has a
long pedigree in applied statistics \cite{Klein-statistical-visions}, and is
prevalent among statisticians, social scientists (especially econometricians)
and engineers.  The younger school, developed essentially since the 1970s,
comes out of physics and nonlinear dynamics.  The first views time series as
samples from a stochastic process, and applies a mixture of traditional
statistical tools and assumptions (linear regression, the properties of
Gaussian distributions) and the analysis of the Fourier spectrum.  The second
school views time series as distorted or noisy measurements of an underlying
dynamical system, which it aims to reconstruct.

The separation between the two schools is in part due to the fact that, when
statistical methods for time series analysis were first being formalized, in
the 1920s and 1930s, dynamical systems theory was literally just beginning.
The real development of nonlinear dynamics into a powerful discipline has
mostly taken place since the 1960s, by which point the statistical theory had
acquired a research agenda with a lot of momentum.  In turn, many of the
physicists involved in experimental nonlinear dynamics in the 1980s and early
1990s were fairly cavalier about statistical issues, and some happily reported
results which should have been left in their file-drawers.

There are welcome signs, however, that the two streams of thought are
coalescing.  Since the 1960s, statisticians have increasingly come to realize
the virtues of what they call ``state-space models'', which are just what the
physicists have in mind with their dynamical systems.  The physicists, in turn,
have become more sensitive to statistical issues, and there is even now some
cross-disciplinary work.  In this section, I will try, so far as possible, to
use the state-space idea as a common framework to present both sets of methods.

\subsection{The State-Space Picture}
\label{sec:state-space-picture}

A vector-valued function of time, $x_t$, the {\bold state}.
In discrete time, this evolves according to some map,
\begin{eqnarray}
x_{t+1} & \equiv & F(x_t,t,\epsilon_t)
\end{eqnarray}
where the map $F$ is allowed to depend on time $t$ and a sequence of
independent random variables $\epsilon_t$.  In continuous time, we do not
specify the evolution of the state directly, but rather the rates of change of
the components of the state,
\begin{eqnarray}
\frac{dx}{dt} &\equiv & F(x,t,\epsilon_t)
\end{eqnarray}
Since our data are generally taken in discrete time, I will restrict myself to
considering that case from now on; almost everything carries over to continuous
time naturally.  The evolution of $x$ is so to speak, self-contained, or more
precisely Markovian: all the information needed to determine the future is
contained in the {\em present} state $x_t$, and earlier states are irrelevant.
(This is basically how physicists {\em define} ``state'' \cite{Dirac-on-qm}.)
Indeed, it is often reasonable to assume that $F$ is independent of time, so
that the dynamics are {\bold autonomous} (in the terminology of dynamics) or
{\bold homogeneous} (in that of statistics).  If we could look at the series of
states, then, we would find it had many properties which made it very
convenient to analyze.

Sadly, however, we do not observe the state $x$; what we observe or measure is
$y$, which is generally a noisy, nonlinear function of the state: $y_t = h(x_t,
\eta_t)$, where $\eta_t$ is measurement noise.  Whether $y$, too, has the
convenient properties depends on $h$, and usually $y$ is {\em not} convenient.
Matters are made more complicated by the fact that we do not, in typical cases,
know the observation function $h$, nor the state-dynamics $F$, nor even,
really, what space $x$ lives in.  The goal of time-series methods is to make
educated guess about all these things, so as to better predict and understand
the evolution of temporal data.

In the ideal case, simply from a knowledge of $y$, we would be able to identify
the state space, the dynamics, and the observation function.  As a matter of
pure mathematical possibility, this can be done for essentially arbitrary
time-series \cite{Knight-predictive-view,Knight-foundations-of-prediction}.
Nobody, however, knows how to do this with complete generality in practice.
Rather, one makes certain assumptions about, say, the state space, which are
strong enough that the remaining details can be filled in using $y$.  Then one
checks the result for accuracy and plausibility, i.e., for the kinds of errors
which would result from breaking those assumptions \cite{Mayo-error}.

Subsequent parts of this section describe classes of such methods.  First,
however, I describe some of the general properties of time series, and
general measurements which can be made upon them.

\paragraph{Notation}

There is no completely uniform notation for time-series.  Since it will be
convenient to refer to sequences of consecutive values. I will write all the
measurements starting at $s$ and ending at $t$ as $y_s^t$.  Further, I will
abbreviate the set of all measurements up to time $t$, $y^{t}_{-\infty}$, as
$y^{-}_{t}$, and the future starting from $t$, $y_{t+1}^{\infty}$, as
$y^{+}_t$.

\subsection{General Properties of Time Series}
\label{sec:time-series-properties}

One of the most commonly assumed properties of a time-series is {\bold
  stationarity}, which comes in two forms: {\bold strong} or {\bold strict}
stationarity, and {\bold weak}, {\bold wide-sense} or {\em second-order}
stationarity.  Strong stationarity is the property that the probability
distribution of sequences of observations does not change over time.  That is,
\begin{eqnarray}
\Prob(Y_{t}^{t+h}) & = & \Prob(Y_{t+\tau}^{t+\tau+h})
\end{eqnarray}
for all lengths of time $h$ and all shifts forwards or backwards in time
$\tau$.  When a series is described as ``stationary'' without qualification, it
depends on context whether strong or weak stationarity is meant.

Weak stationarity, on the other hand, is the property that the first and second
moments of the distribution do not change over time.
\begin{eqnarray}
\Expec{Y_t} & = & \Expec{Y_{t+\tau}}\\
\Expec{Y_t Y_{t+h}} & = & \Expec{Y_{t+\tau} Y_{t+\tau+h}}
\end{eqnarray}
If $Y$ is a Gaussian process, then the two senses of stationarity are
equivalent.  Note that both sorts of stationarity are statements about the true
distribution, and so cannot be simply read off from measurements.

Strong stationarity implies a property called {\bold ergodicity}, which is much
more generally applicable.  Roughly speaking, a series is ergodic if any
sufficiently long sample is representative of the entire process.  More
exactly, consider the {\bold time-average} of a well-behaved function $f$ of
$Y$,
\begin{eqnarray}
{\TimeAvg{f}}_{t_1}^{t_2} & \equiv & \frac{1}{t_2 - t_1}\sum_{t = t_1}^{t = t_2}{f(Y_t)} ~.
\end{eqnarray}
This is generally a random quantity, since it depends on where the trajectory
started at $t_1$, and any random motion which may have taken place between then
and $t_2$.  Its distribution generally depends on the precise values of $t_1$
and $t_2$.  The series $Y$ is ergodic if almost all time-averages converge
eventually, i.e., if
\begin{eqnarray}
\lim_{T \rightarrow \infty}{{\TimeAvg{f}}_{t}^{t+T}} & = & \bar{f}
\end{eqnarray}
for some constant $\bar{f}$ independent of the starting time $t$, the starting
point $Y_t$, or the trajectory $Y^{\infty}_{t}$.  {\bold Ergodic theorems}
specify conditions under which ergodicity holds; surprisingly, even completely
deterministic dynamical systems can be ergodic.

Ergodicity is such an important property because it means that statistical
methods are very directly applicable.  Simply by waiting long enough, one can
obtain an estimate of any desired property which will be closely representative
of the future of the process.  Statistical inference {\em is} possible for
non-ergodic processes, but it is considerably more difficult, and often
requires multiple time-series
\cite{Gray-ergodic-properties,Basawa-Scott-non-ergodic}.

One of the most basic means of studying a time series is to compute the {\bold
  autocorrelation function} (ACF), which measures the linear dependence between
the values of the series at different points in time.  This starts with the
{\bold autocovariance function}:
\begin{eqnarray}
C(s,t) & \equiv & \Expec{\left(y_s - \Expec{y_s}\right)\left(y_t - \Expec{y_t}\right)} ~.
\end{eqnarray}
(Statistical physicists, unlike anyone else, call {\em this} the ``correlation
function''.)  The autocorrelation itself is the autocovariance, normalized by
the variability of the series:
\begin{eqnarray}
\rho(s,t) & \equiv & \frac{C(s,t)}{\sqrt{C(s,s) C(t,t)}}
\end{eqnarray}
$\rho$ is $\pm 1$ when $y_s$ is a linear function of $y_t$.  Note that the
definition is symmetric, so $\rho(s,t) = \rho(t,s)$.  For stationary or
weakly-stationary processes, one can show that $\rho$ depends only on the
difference between $\tau$ $t$ and $s$.  In this case one just writes
$\rho(\tau)$, with one argument.  $\rho(0) = 1$, always.  The time $t_c$ such
that $\rho(t_c) = 1/e$ is called the {\bold (auto)correlation time} of the
series.

The correlation function is a {\bold time-domain} property, since it is
basically about the series considered as a sequence of values at distinct
times.  There are also {\bold frequency-domain} properties, which depend on
re-expressing the series as a sum of sines and cosines with definite
frequencies.  A function of time $y$ has a Fourier transform which is a
function of frequency, $\tilde{y}$.
\begin{eqnarray}
\tilde{y} & \equiv & \mathcal{F}y\\
\tilde{y}_{\nu} &=& \sum_{t=1}^T e^{-i\frac{2\pi \nu t}{T}} y_t ~,
\end{eqnarray}
assuming the time series runs from $t=1$ to $t=T$.  (Rather than separating out
the sine and cosine terms, it is easier to use the complex-number
representation, via $e^{i\theta} = \cos{\theta} + i\sin{\theta}$.)  The inverse
Fourier transform recovers the original function:
\begin{eqnarray}
y & = & \mathcal{F}^{-1} \tilde{y}\\
y_t &=& \frac{1}{T} \sum_{\nu=0}^{T-1} e^{i\frac{2\pi \nu t}{T}} \tilde{y}_{\nu}
\end{eqnarray}
The Fourier transform is a linear operator, in the sense that $\mathcal{F}(x +
y) = \mathcal{F}x + \mathcal{F}y$.  Moreover, it represents series we are
interested in as a sum of trigonometric functions, which are themselves
solutions to linear differential equations.  These facts lead to extremely
powerful frequency-domain techniques for studying linear systems.  Of course,
the Fourier transform is always {\em valid}, whether the system concerned is
linear or not, and it may well be useful, though that is not guaranteed.

The squared absolute value of the Fourier transform, $f(\nu) =
{|\tilde{y_{\nu}}|}^2$, is called the {\bold spectral density} or {\bold power
  spectrum}.  For stationary processes, the power spectrum $f(\nu)$ is the
Fourier transform of the autocovariance function $C(\tau)$ (a result called the
Wiener-Khinchin theorem).  An important consequence is that a Gaussian process
is completely specified by its power spectrum.  In particular, consider a
sequence of independent Gaussian variables, each with variance $\sigma^2$.
Because they are perfectly uncorrelated, $C(0) = \sigma^2$, and $C(\tau) = 0$
for any $\tau \neq 0$.  The Fourier transform of such a $C(\tau)$ is just
$f(\nu) = \sigma^2$, independent of $\nu$ --- every frequency has just as much
power.  Because white light has equal power in every color of the spectrum,
such a process is called {\bold white noise}.  Correlated processes, with
uneven power spectra, are sometimes called {\bold colored noise}, and there is
an elaborate terminology of red, pink, brown, etc. noises
\cite[ch.\ 3]{West-Deering-lure}.

The easiest way to estimate the power spectrum is simply to take the Fourier
transform of the time series, using, e.g., the fast Fourier transform algorithm
\cite{Numerical-Recipes-in-C}.  Equivalently, one might calculate the
autocovariance and Fourier transform that.  Either way, one has an estimate of
the spectrum which is called the {\bold periodogram}.  It is unbiased, in that
the expected value of the periodogram at a given frequency is the true power at
that frequency.  Unfortunately, it is not consistent --- the variance around
the true value does not shrink as the series grows.  The easiest way to
overcome this is to apply any of several well-known smoothing functions to the
periodogram, a procedure called {\bold windowing} \cite{Shumway-Stoffer}.
(Standard software packages will accomplish this automatically.)

The Fourier transform takes the original series and decomposes it into a sum of
sines and cosines.  This is possible because {\em any} reasonable function can
be represented in this way.  The trigonometric functions are thus a {\bold
  basis} for the space of functions.  There are many other possible bases, and
one can equally well perform the same kind of decomposition in any other basis.
The trigonometric basis is particularly useful for stationary time series
because the basis functions are themselves evenly spread over all times
\cite[ch.\ 2]{Wiener-cybernetics}.  Other bases, localized in time, are more
convenient for non-stationary situations.  The most well-known of these
alternate bases, currently, are wavelets \cite{World-according-to-wavelets},
but there is, literally, no counting the other possibilities.

\subsection{The Traditional Statistical Approach}

The traditional statistical approach to time series is to represent them
through linear models of the kind familiar from applied statistics.

The most basic kind of model is that of a {\bold moving average}, which is
especially appropriate if $x$ is highly correlated up to some lag, say $q$,
after which the ACF decays rapidly.  The moving average model represents $x$ as
the result of smoothing $q+1$ independent random variables.  Specifically, the
MA($q$) model of a weakly stationary series is
\begin{eqnarray}
y_t & = & \mu + w_t + \sum_{k=1}^{q}{\theta_k w_{t-k}}
\end{eqnarray}
where $\mu$ is the mean of $y$, the $\theta_i$ are constants and the $w_t$ are
white noise variables.  $q$ is called the {\bold order} of the model.  Note
that there is no direct dependence between successive values of $y$; they are
all functions of the white noise series $w$.  Note also that $y_t$ and
$y_{t+q+1}$ are completely independent; after $q$ time-steps, the effects of
what happened at time $t$ disappear.

Another basic model is that of an {\bold autoregressive process}, where the
next value of $y$ is a linear combination of the preceding values of $y$.
Specifically, an AR($p$) model is
\begin{eqnarray}
y_t & = & \alpha + \sum_{k=1}^{p}{\phi_k y_{t-k}} + w_t
\end{eqnarray}
where $\phi_i$ are constants and $\alpha = \mu(1 - \sum_{k=1}^{p}{\phi_k})$.
The order of the model, again is $p$.  This is the multiple regression of
applied statistics transposed directly on to time series, and is surprisingly
effective.  Here, unlike the moving average case, effects propagate
indefinitely --- changing $y_t$ can affect all subsequent values of $y$.  The
remote past only becomes irrelevant if one controls for the last $p$ values of
the series.  If the noise term $w_t$ were absent, an AR($p$) model would be a
$p^{\mathrm th}$ order linear difference equation, the solution to which would
be some combination of exponential growth, exponential decay and harmonic
oscillation.  With noise, they become oscillators under stochastic forcing
\cite{Honerkamp-stochastic}.

The natural combination of the two types of model is the {\bold autoregressive
  moving average model}, ARMA($p,q$):
\begin{eqnarray}
y_t & = & \alpha + \sum_{k=1}^{p}{\phi_k y_{t-k}} + w_t + \sum_{k=1}^{q}{\theta_k w_{t-k}}
\end{eqnarray}
This combines the oscillations of the AR models with the correlated driving
noise of the MA models.  An AR($p$) model is the same as an ARMA($p,0$) model,
and likewise an MA($q$) model is an ARMA($0,q$) model.

It is convenient, at this point in our exposition, to introduce the notion of
the {\bold back-shift operator} $B$,
\begin{eqnarray}
\label{backshift}
By_t & = & y_{t-1} ~,
\end{eqnarray}
and the {\bold AR and MA polynomials},
\begin{eqnarray}
\phi(z) & = & 1 - \sum_{k=1}^{p}{\phi_k z^k} ~,\\
\theta(z) & = & 1 + \sum_{k=1}^{q}{\theta_k z^k} ~,
\end{eqnarray}
respectively.  Then, formally speaking, in an ARMA process is
\begin{eqnarray}
\phi(B)y_t  & = & \theta(B) w_t ~.
\end{eqnarray}
The advantage of doing this is that one can determine many properties of an
ARMA process by algebra on the polynomials.  For instance, two important
properties we want a model to have are {\bold invertibility} and {\bold
  causality}.  We say that the model is invertible if the sequence of noise
variables $w_t$ can be determined uniquely from the observations $y_t$; in this
case we can write it as an MA($\infty$) model.  This is possible just when
$\theta(z)$ has no roots inside the unit circle.  Similarly, we say the model
is causal if it can be written as an AR($\infty$) model, without reference to
any {\em future} values.  When this is true, $\phi(z)$ also has no roots inside
the unit circle.

If we have a causal, invertible ARMA model, with known parameters, we can work
out the sequence of noise terms, or {\bold innovations} $w_t$ associated with
our measured values $y_t$.  Then, if we want to forecast what happens past the
end of our series, we can simply extrapolate forward, getting predictions
$\hat{y}_{T+1}, \hat{y}_{T+2}$, etc.  Conversely, if we knew the innovation
sequence, we could determine the parameters $\phi$ and $\theta$.  When both
are unknown, as is the case when we want to fit a model, we need to determine
them jointly \cite{Shumway-Stoffer}.  In particular, a common procedure is to
work forward through the data, trying to predict the value at each time on the
basis of the past of the series; the sum of the squared differences between
these predicted values $\hat{y}_t$ and the actual ones $y_t$ forms the
empirical loss:
\begin{eqnarray}
L & = & \sum_{i=1}^{T}{{(y_t - \hat{y}_t)}^2}
\end{eqnarray}
For this loss function, in particular, there are very fast standard algorithms,
and the estimates of $\phi$ and $\theta$ converge on their true values,
provided one has the right model order.

This leads naturally to the question of how one determines the order of ARMA
model to use, i.e., how one picks $p$ and $q$.  This is precisely a model
selection task, as discussed in \S \ref{sec:data-mining}.  All methods
described there are potentially applicable; cross-validation and regularization
are more commonly used than capacity control.  Many software packages will
easily implement selection according to the AIC, for instance.

The power spectrum of an ARMA($p,q$) process can be given in closed form:
\begin{eqnarray}
f(\nu) & = &
\frac{\sigma^2}{2\pi}\frac{{(1+\sum_{k=1}^{q}{\theta_k e^{-i\nu k}})}^2}{{(1-\sum_{k=1}^{p}{\phi_k e^{-i\nu k}})}^2}~.
\end{eqnarray}
Thus, the parameters of an ARMA process can be estimated directly from the
power spectrum, if you have a reliable estimate of the spectrum.  Conversely,
different hypotheses about the parameters can be checked from spectral data.

All ARMA models are weakly stationary; to apply them to non-stationary data one
must transform the data so as to make it stationary.  A common transformation
is {\bold differencing}, i.e., applying operations of the form
\begin{eqnarray}
\nabla y_t & = & y_t - y_{t-1} ~,
\end{eqnarray}
which tends to eliminate regular trends.  In terms of the back-shift operator,
\begin{eqnarray}
\nabla y_t & = & (1 - B)y_t ~,
\end{eqnarray}
and higher-order differences are 
\begin{eqnarray}
\nabla^d y_t & = & (1-B)^d y_t ~.
\end{eqnarray}
Having differenced the data to our satisfaction, say $d$ times, we then fit an
ARMA model to it.  The result is an {\bold autoregressive integrated moving
  average} model, ARIMA($p,d,q$) \cite{Box-Jenkins}, given by
\begin{eqnarray}
\phi(B)(1-B)^dy_t = \theta(B)w_t ~,
\end{eqnarray}

As mentioned above (\S \ref{sec:state-space-picture}), ARMA and ARIMA models
can be re-cast in state space terms, so that our $y$ is a noisy measurement of
a hidden $x$ \cite{Durbin-Koopman-state-space-methods}.  For these models, both
the dynamics and the observation functions are linear, that is, $x_{t+1} =
\mathbf{A}x_t + \epsilon_t$ and $y_{t} = \mathbf{B}x_t + \eta_t$, for some
matrices $\mathbf{A}$ and $\mathbf{B}$.  The matrices can be determined from
the $\theta$ and $\phi$ parameters, though the relation is a bit too involved
to give here.

\subsubsection{Applicability of Linear Statistical Models}

It is often possible to describe a nonlinear dynamical system through an
effective linear statistical model, provided the nonlinearities are cooperative
enough to appear as noise \cite{Eyink-linear-stochastic-models}.  It is an
under-appreciated fact that this is at least sometimes true even of turbulent
flows \cite{Barndorff-Nielsen-et-al-parametric-turbulence,%
  Eyink-Alexander-predictive-turbulence}; the generality of such an approach is
not known.  Certainly, if you care only about predicting a time series, and not
about its structure, it is always a good idea to try a linear model first, even
if you {\em know} that the real dynamics are highly nonlinear.

\subsubsection{Extensions}

While standard linear models are more flexible than one might think, they do
have their limits, and recognition of this has spurred work on many extensions
and variants.  Here I briefly discuss a few of these.

\paragraph{Long Memory}
The correlations of standard ARMA and ARIMA models decay fairly rapidly, in
general exponentially; $\rho(t) \propto e^{-t/\tau_c}$, where $\tau_c$ is the
correlation time.  For some series, however, $\tau_c$ is effectively infinite,
and $\rho(t) \propto t^{-\alpha}$ for some exponent $\alpha$.  These are called
{\bold long-memory processes}, because they remain substantially correlated
over very long times.  These can still be accommodated within the ARIMA
framework, formally, by introducing the idea of {\em fractional} differencing,
or, in continuous time, fractional derivatives
\cite{Beran-long-memory,West-Deering-lure}.  Often long-memory processes are
self-similar, which can simplify their statistical estimation
\cite{Embrechts-Maejima-book}.

\paragraph{Volatility}
All ARMA and even ARIMA models assume constant variance.  If the variance is
itself variable, it can be worthwhile to model it.  {\bold Autoregressive
  conditionally heteroscedastic} (ARCH) models assume a fixed mean value for
$y_r$, but a variance which is an auto-regression on $y_t^2$.  {\bold
  Generalized ARCH} (GARCH) models expand the regression to include the
(unobserved) earlier variances.  ARCH and GARCH models are especially suitable
for processes which display {\bold clustered volatility}, periods of extreme
fluctuation separated by stretches of comparative calm.

\paragraph{Nonlinear and Nonparametric Models}
Nonlinear models are obviously appealing, and when a particular parametric form
of model is available, reasonably straight-forward modifications of the linear
machinery can be used to fit, evaluate and forecast the model \cite[\S
  4.9]{Shumway-Stoffer}.  However, it is often impractical to settle on a good
parametric form beforehand.  In these cases, one must turn to nonparametric
models, as discussed in \S \ref{sec:architectures}; neural networks are a
particular favorite here \cite{Zapranis-Refenes}.  The so-called {\bold kernel
  smoothing methods} are also particularly well-developed for time series, and
often perform almost as well as parametric models \cite{Bosq-nonparametric}.
Finally, information theory provides {\bold universal prediction methods},
which promise to asymptotically approach the best possible prediction, starting
from exactly no background knowledge.  This power is paid for by demanding a
long initial training phase used to infer the structure of the process, when
predictions are much worse than many other methods could deliver
\cite{Algoet-universal-schemes}.

\subsection{The Nonlinear Dynamics Approach}
\label{sec:nld-approach}

The younger approach to the analysis of time series comes from nonlinear
dynamics, and is intimately bound up with the state-space approach described in
\S \ref{sec:state-space-picture} above.  The idea is that the dynamics on the
state space can be determined {\em directly} from observations, at least if
certain conditions are met.

The central result here is the Takens Embedding Theorem
\cite{Takens-embedding}; a simplified, slightly inaccurate version is as
follows.  Suppose the $d$-dimensional state vector $x_t$ evolves according to
an unknown but continuous and (crucially) deterministic dynamic.  Suppose, too,
that the one-dimensional observable $y$ is a smooth function of $x$, and
``coupled'' to all the components of $x$.  Now at any time we can look not just
at the present measurement $y(t)$, but also at observations made at times
removed from us by multiples of some lag $\tau$: $y_{t-\tau}$, $y_{t-2\tau}$,
etc.  If we use $k$ lags, we have a $k$-dimensional vector.  One might expect
that, as the number of lags is increased, the motion in the lagged space will
become more and more predictable, and perhaps in the limit $k\rightarrow\infty$
would become deterministic.  In fact, the dynamics of the lagged vectors become
deterministic at a finite dimension; not only that, but the deterministic
dynamics are completely equivalent to those of the original state space!  (More
exactly, they are related by a smooth, invertible change of coordinates, or
{\bold diffeomorphism}.)  The magic {\bold embedding dimension} $k$ is at most
$2d+1$, and often less.

Given an appropriate reconstruction via embedding, one can investigate many
aspects of the dynamics.  Because the reconstructed space is related to the
original state space by a smooth change of coordinates, any geometric property
which survives such treatment is the same for both spaces.  These include the
dimension of the attractor, the Lyapunov exponents (which measure the degree of
sensitivity to initial conditions) and certain qualitative properties of the
autocorrelation function and power spectrum (``correlation dimension'').  Also
preserved is the relation of ``closeness'' among trajectories --- two
trajectories which are close in the state space will be close in the embedding
space, and vice versa.  This leads to a popular and robust scheme for nonlinear
prediction, the {\bold method of analogs}: when one wants to predict the next
step of the series, take the current point in the embedding space, find a
similar one with a known successor, and predict that the current point will do
the analogous thing.  Many refinements are possible, such as taking a weighted
average of nearest neighbors, or selecting an analog at random, with a
probability decreasing rapidly with distance.  Alternately, one can simply fit
non-parametric predictors on the embedding space.  (See \cite{Kantz-Schreiber}
for a review.)  Closely related is the idea of {\bold noise reduction}, using
the structure of the embedding-space to filter out some of the effects of
measurement noise.  This can work even when the statistical character of the
noise is unknown (see \cite{Kantz-Schreiber} again).

Determining the number of lags, and the lag itself, is a problem of model
selection, just as in \S \ref{sec:data-mining}, and can be approached in that
spirit.  An obvious approach is to minimize the in-sample forecasting error, as
with ARMA models; recent work along these lines
\cite{Judd-Mees-embedding-as-modeling,Small-Tse-optimal-embedding} uses the
minimum description length principle (described in \S \ref{sec:mdl} below) to
control over-fitting.  A more common procedure for determining the embedding
dimension, however, is the {\bold false nearest neighbor method}
\cite{Kennel-Brown-Abarbanel-false-nearest-neighbors}.  The idea is that if the
current embedding dimension $k$ is sufficient to resolve the dynamics, $k+1$
would be too, and the reconstructed state space will not change very much.  In
particular, points which were close together in the dimension-$k$ embedding
should remain close in the dimension-$k+1$ embedding.  Conversely, if the
embedding dimension is too small, points which are really far apart will be
brought artificially close together (just as projecting a sphere on to a disk
brings together points on the opposite side of a sphere).  The particular
algorithm of Kennel et al., which has proved very practical, is to take each
point in the $k$-dimensional embedding, find its nearest neighbor in that
embedding, and then calculate the distance between them.  One then calculates
how much further apart they would be if one used a $k+1$-dimensional embedding.
If this extra distance is more than a certain fixed multiple of the original
distance, they are said to be ``false nearest neighbors''.  (Ratios of 2 to 15
are common, but the precise value does not seem to matter very much.)  One then
repeats the process at dimension $k+1$, stopping when the proportion of false
nearest neighbors becomes zero, or at any rate sufficiently small.  Here, the
loss function used to guide model selection is the number of false nearest
neighbors, and the standard prescriptions amount to empirical risk
minimization.  One reason simple ERM works well here is that the problem is
intrinsically finite-dimensional (via the Takens result).

Unfortunately, the data required for calculations of quantities like dimensions
and exponents to be reliable can be quite voluminous.  Approximately
${10}^{2+0.4D}$ data-points are necessary to adequately reconstruct an
attractor of dimension $D$ \cite[pp.\ 317--319]{Sprott-on-time-series}.  (Even
this is more optimistic than the widely-quoted, if apparently pessimistic,
calculation of \cite{Smith-on-embedding-dimension}, that attractor
reconstruction with an {\em embedding} dimension of $k$ needs ${42}^{k}$
data-points!)  In the early days of the application of embedding methods to
experimental data, these limitations were not well appreciated, leading to many
calculations of low-dimensional deterministic chaos in EEG and EKG series,
economic time series, etc., which did not stand up to further scrutiny.  This
in turn brought some discredit on the methods themselves, which was not really
fair.  More positively, it also led to the development of ideas such as {\bold
  surrogate-data methods}.  Suppose you have found what seems like a good
embedding, and it appears that your series was produced by an underlying
deterministic attractor of dimension $D$.  One way to test this hypothesis
would be to see what kind of results your embedding method would give if
applied to similar but {\em non}-deterministic data.  Concretely, you find a
stochastic model with similar statistical properties (e.g., an ARMA model with
the same power spectrum), and simulate many time-series from this model.  You
apply your embedding method to each of these {\bold surrogate data} series,
getting the approximate distribution of apparent ``attractor'' dimensions when
there really is no attractor.  If the dimension measured from the original data
is not significantly different from what one would expect under this null
hypothesis, the evidence for an attractor (at least from this source) is weak.
To apply surrogate data tests well, one must be very careful in constructing
the null model, as it is easy to use over-simple null models, biasing the test
towards apparent determinism.

A few further cautions on embedding methods are in order.  While {\em in
  principle} any lag $\tau$ is suitable, in practice very long or very short
lags both lead to pathologies.  A common practice is to set the lag to the
autocorrelation time (see above), or the first minimum of the mutual
information function (see \S \ref{sec:info-theory} below), the notion being
that this most nearly achieves a genuinely ``new'' measurement
\cite{Fraser-Swinney-independent-coords}.  There is some evidence that the
mutual information method works better
\cite{Cellucci-et-al-comparative-embedding-methods}.  Again, while in principle
almost any smooth observation function will do, given enough data, in practice
some make it much easier to reconstruct the dynamics; several {\bold indices of
  observability} try to quantify this
\cite{Letellier-Aguirre-symmetries-and-observables}.  Finally, it strictly
applies only to deterministic observations of deterministic systems.  Embedding
approaches are reasonably robust to a degree of noise in the observations.
They do not cope at all well, however, to noise in the dynamics itself.  To
anthropomorphize a little, when confronted by apparent non-determinism, they
respond by adding more dimensions, and so distinguishing apparently similar
cases.  Thus, when confronted with data which really are stochastic, they will
infer an infinite number of dimensions, which is correct in a way, but
definitely not helpful.

These remarks should not be taken to belittle the very real power of nonlinear
dynamics methods.  Applied skillfully, they are powerful tools for
understanding the behavior of complex systems, especially for probing aspects
of their structure which are not directly accessible.

\subsection{Filtering and State Estimation}
\label{sec:filtering}

Suppose we have a state-space model for our time series, and some observations
$y$, can we find the state $x$?  This is the problem of {\bold filtering} or
{\bold state estimation}.  Clearly, it is not the same as the problem of
finding a model in the first place, but it is closely related, and also a
problem in statistical inference.

In this context, a {\bold filter} is a function which provides an estimate
$\hat{x}_t$ of $x_t$ on the basis of observations up to and
including\footnote{One could, of course, build a filter which uses later $y$
  values as well; this is a {\bold non-causal} or {\bold smoothing} filter.
  This is clearly not suitable for estimating the state in real time, but often
  gives more accurate estimates when it is applicable.  The discussion in the
  text generally applies to smoothing filters, at some cost in extra notation.}
time $t$: $\hat{x}_t = f(y_0^t)$.  A filter is {\bold
  recursive}\footnote{Equivalent terms are {\bold future-resolving} or {\bold
    right-resolving} (from nonlinear dynamics) and {\bold deterministic} (the
  highly confusing contribution of automata theory).}  if it estimates the
state at $t$ on the basis of its estimate at $t-1$ and the new observation:
$\hat{x}_t = f(\hat{x}_{t-1}, y_t)$.  Recursive filters are especially suited
to on-line use, since one does not need to retain the complete sequence of
previous observations, merely the most recent estimate of the state.  As with
prediction in general, filters can be designed to provide either point
estimates of the state, or distributional estimates.  Ideally, in the latter
case, we would get the conditional distribution, $\Prob(X_t = x|Y_1^t =
y_1^t)$, and in the former case the conditional expectation,
$\int_{x}{x\Prob(X_t = x|Y_1^t = y_1^t) dx}$.

Given the frequency with which the problem of state estimation shows up in
different disciplines, and its general importance when it does appear, much
thought has been devoted to it over many years.  The problem of optimal {\em
  linear} filters for stationary processes was solved independently by two of
the ``grandfathers'' of complex systems science, Norbert Wiener and
A. N. Kolmogorov, during the Second World War
\cite{Wiener-time-series,Kolmogorov-interpolation-extrapolation}.  In the
1960s, Kalman and Bucy \cite{Kalman,Kalman-Bucy,Bucy-filtering} solved the
problem of optimal recursive filtering, assuming linear dynamics, linear
observations and additive noise.  In the resulting {\bold Kalman filter}, the
new estimate of the state is a weighted combination of the old state,
extrapolated forward, and the state which would be inferred from the new
observation alone.  The requirement of linear dynamics can be relaxed slightly
with what's called the ``extended Kalman filter'', essentially by linearizing
the dynamics around the current estimated state.

Nonlinear solutions go back to pioneering work of Stratonovich
\cite{Stratonovich-conditional-markov-processes} and Kushner
\cite{Kushner-optimal-nonlinear-filtering} in the later 1960s, who gave
optimal, recursive solutions.  Unlike the Wiener or Kalman filters, which give
point estimates, the Stratonovich-Kushner approach calculates the complete
conditional distribution of the state; point estimates take the form of the
mean or the most probable state \cite{Lipster-Shiryaev}.  In most
circumstances, the strictly optimal filter is hopelessly impractical
numerically.  Modern developments, however, have opened up some very important
lines of approach to practical nonlinear filters \cite{Tanizaki-filters},
including approaches which exploit the geometry of the nonlinear dynamics
\cite{Darling-geometrically-intrinsic-filters-1,%
  Darling-geometrically-intrinsic-filters-2}, as well as more mundane methods
which yield tractable numerical approximations to the optimal filters
\cite{Eyink-variational-optimal-estimation,Ahmed-filtering}.  Noise reduction
methods (\S \ref{sec:nld-approach}) and hidden Markov models (\S
\ref{sec:symbolic-dyn}) can also be regarded as nonlinear filters.

\subsection{Symbolic or Categorical Time Series}
\label{sec:symbolic-dyn}

The methods we have considered so far are intended for time-series taking
continuous values.  An alternative is to break the range of the time-series
into discrete categories (generally only finitely many of them); these
categories are sometimes called {\bold symbols}, and the study of these
time-series {\bold symbolic dynamics}.  Modeling and prediction then reduces to
a (perhaps more tractable) problem in discrete probability, and many methods
can be used which are simply inapplicable to continuous-valued series
\cite{Badii-Politi}.  Of course, if a bad discretization is chosen, the results
of such methods are pretty well meaningless, but sometimes one gets data which
is already nicely discrete --- human languages, the sequences of bio-polymers,
neuronal spike trains, etc.  We shall return to the issue of discretization
below, but for the moment, we will simply consider the applicable methods for
discrete-valued, discrete-time series, however obtained.

Formally, we take a continuous variable $z$ and {\bold partition} its range
into a number of discrete {\bold cells}, each labeled by a different symbol
from some {\bold alphabet}; the partition gives us a discrete variable $y =
\phi(z)$.  A {\bold word} or {\bold string} is just a sequence of symbols, $y_0
y_1 \ldots y_n$.  A time series $z_0^n$ naturally generates a string
$\phi(z_0^n) \equiv \phi(z_0) \phi(z_1) \ldots \phi(z_n)$.  In general, not
every possible string can actually be generated by the dynamics of the system
we're considering.  The set of allowed sequences is called the {\bold
  language}.  A sequence which is never generated is said to be {\bold
  forbidden}.  In a slightly inconsistent metaphor, the rules which specify the
allowed words of a language are called its {\bold grammar}.  To each grammar
there corresponds an abstract machine or {\bold automaton} which can determine
whether a given word belongs to the language, or, equivalently, generate all
and only the allowed words of the language.  The generative versions of these
automata are stochastic, i.e., they generate different words with different
probabilities, matching the statistics of $\phi(z)$.

By imposing restrictions on the forms the grammatical rules can take, or,
equivalently, on the memory available to the automaton, we can divide all
languages into four nested classes, a hierarchical classification due to
Chomsky \cite{Chomsky-three-models}.  At the bottom are the members of the
weakest, most restricted class, the {\bold regular languages} generated by
automata within only a fixed, finite memory for past symbols ({\bold finite
  state machines}).  Above them are the {\bold context free} languages, whose
grammars do not depend on context; the corresponding machines are {\bold stack
  automata}, which can store an unlimited number of symbols in their memory,
but on a strictly first-in, first-out basis.  Then come the {\bold
  context-sensitive} languages; and at the very top, the unrestricted
languages, generated by universal computers.  Each stage in the hierarchy can
simulate all those beneath it.

We may seem to have departed very far from dynamics, but actually this is not
so.  Because different languages classes are distinguished by different kinds
of memories, they have very different correlation properties (\S
\ref{sec:time-series-properties}), mutual information functions (\S
\ref{sec:info-theory}), and so forth --- see \cite{Badii-Politi} for details.
Moreover, it is often easier to determine these properties from a system's
grammar than from direct examination of sequence statistics, especially since
specialized techniques are available for grammatical inference
\cite{Charniak,Manning-Schutze}.

\subsubsection{Hidden Markov Models}

The most important special case of this general picture is that of regular
languages.  These, we said, are generated by machines with only a finite
memory.  More exactly, there is a finite set of states $x$, with two
properties:
\begin{enumerate}
\item The distribution of $y_t$ depends solely on $x_t$, and
\item The distribution of $x_{t+1}$ depends solely on $x_t$.
\end{enumerate}
That is, the $x$ sequence is a Markov chain, and the observed $y$ sequence is
noisy function of that chain.  Such models are very familiar in signal
processing \cite{Elliott-et-al-HMM}, bioinformatics \cite{Baldi-Brunak-bioinfo}
and elsewhere, under the name of {\bold hidden Markov models} (HMMs).  They can
be thought of as a generalization of ordinary Markov chains to the state-space
picture described in \S \ref{sec:state-space-picture}.  HMMs are particularly
useful in filtering applications, since very efficient algorithms exist for
determining the most probable values of $x$ from the observed sequence $y$.
The {\bold expectation-maximization} (EM) algorithm
\cite{Neal-Hinton-view-of-EM} even allows us to simultaneously infer the most
probable hidden states and the most probable parameters for the model.

\subsubsection{Variable-Length Markov Models}
\label{sec:VLMMs}

The main limitation of ordinary HMMs methods, even the EM algorithm, is that
they assume a fixed {\bold architecture} for the states, and a fixed
relationship between the states and the observations.  That is to say, they are
not geared towards inferring the structure of the model.  One could apply the
model-selection techniques of \S \ref{sec:data-mining}, but methods of direct
inference have also been developed.  A popular one relies on {\bold
  variable-length Markov models}, also called {\bold context trees} or {\bold
  probabilistic suffix trees}
\cite{Rissanen-1983,Willems-Shtarkov-Tjalkens-CTW,Ron-Singer-Tishby-amnesia,%
  Buhlmann-Wyner}.

A suffix here is the string at the end of the $y$ time series at a given time,
so e.g. the binary series $abbabbabb$ has suffixes $b$, $bb$, $abb$, $babb$,
etc., but not $bab$.  A suffix is a {\bold context} if the future of the series
is independent of its past, given the suffix.  Context-tree algorithms try to
identify contexts by iteratively considering longer and longer suffixes, until
they find one which seems to be a context.  For instance, in a binary series,
such an algorithm would first try whether the suffices $a$ and $b$ are
contexts, i.e., whether the conditional distribution $\Prob(Y_{t+1}|Y_t = a)$
can be distinguished from $\Prob(Y_{t+1}|Y_t = a, Y^{-}_{t-1})$, and likewise
for $Y_t = b$. It could happen that $a$ is a context but $b$ is not, in which
case the algorithm will try $ab$ and $bb$, and so on.  If one sets $x_t$ equal
to the context at time $t$, $x_t$ is a Markov chain.  This is called a {\em
  variable-length} Markov model because the contexts can be of different
lengths.

Once a set of contexts has been found, they can be used for prediction.  Each
context corresponds to a different distribution for one-step-ahead predictions,
and so one just needs to find the context of the current time series.  One
could apply state-estimation techniques to find the context, but an easier
solution is to use the construction process of the contexts to build a decision
tree (\S \ref{sec:architectures}), where the first level looks at $Y_t$, the
second at $Y_{t-1}$, and so forth.

Variable-length Markov models are conceptually simple, flexible, fast, and
frequently more accurate than other ways of approaching the symbolic dynamics
of experimental systems \cite{Kennel-Mees-context-tree-modeling}.  However, not
every regular language can be represented by a finite number of contexts.  This
weakness can be remedied by moving to a more powerful class of models,
discussed next.

\subsubsection{Causal-State Models, Observable-Operator Models, and
  Predictive-State Representations}
\label{sec:causal-state-models}

In discussing the state-space picture in \S \ref{sec:state-space-picture}
above, we saw that the state of a system is basically defined by specifying its
future time-evolution, to the extent that it can be specified.  Viewed in this
way, a state $X_t$ corresponds to a distribution over future observables
$Y_{t+1}^{+}$.  One natural way of finding such distributions is to look at the
{\em conditional} distribution of the future observations, given the previous
history, i.e., $\Prob(Y_{t+1}^{+}|Y_t^{-} = y_t^{-})$.  For a given stochastic
process or dynamical system, there will be a certain characteristic family of
such conditional distributions.  One can then consider the distribution-valued
process generated by the original, observed process.  It turns out that the
former has is always a Markov process, and that the original process can be
expressed as a function of this Markov process plus noise.  In fact, the
distribution-valued process has all the properties one would want of a
state-space model of the observations.  The conditional distributions, then,
can be treated as states.

This remarkable fact has lead to techniques for modeling discrete-valued time
series, all of which attempt to capture the conditional-distribution states,
and all of which are strictly more powerful than VLMMs.  There are at least
three: the {\bold causal-state models} or {\bold causal-state machines}
(CSMs)\footnote{Early publications on this work started with the assumption
  that the discrete values were obtained by dividing continuous measurements
  into bins of width $\epsilon$, and so called the resulting models
  ``$\epsilon$-machines''.  This name is unfortunate: that is usually a bad way
  of discretizing data (\S \ref{sec:generating-partitions}), the quantity
  $\epsilon$ plays no role in the actual theory, and the name is more than
  usually impenetrable to outsiders.  While I have used it extensively myself,
  it should probably be avoided.}  introduced by Crutchfield and Young
\cite{Inferring-stat-compl}, the {\bold observable operator models} (OOMs)
introduced by Jaeger \cite{Jaeger-operator-models}, and the {\bold predictive
  state representations} (PSRs) introduced by Littman, Sutton and Singh
\cite{predictive-representations-of-state}.  The simplest way of thinking of
such objects is that they are VLMMs where a context or state can contain more
than one suffix, adding expressive power and allowing them to give compact
representations of a wider range of processes.  (See \cite{CSSR-for-UAI} for
more on this point, with examples.)

All three techniques --- CSMs, OOMs and PSRs --- are basically equivalent,
though they differ in their formalisms and their emphases.  CSMs focus on
representing states as classes of histories with the same conditional
distributions, i.e., as suffixes sharing a single context.  (They also feature
in the ``statistical forecasting'' approach to measuring complexity, discussed
in \S \ref{sec:comp-mech} below.)  OOMs are named after the operators which
update the state; there is one such operator for each possible observation.
PSRs, finally, emphasize the fact that one does not actually need to know the
probability of every possible string of future observations, but just a
restricted sub-set of key trajectories, called ``tests''.  In point of fact,
all of them can be regarded as special cases of more general prior
constructions due to Salmon (``statistical relevance basis'')
\cite{Salmon-1971,Salmon-1984} and Knight (``measure-theoretic prediction
process'') \cite{Knight-predictive-view,Knight-foundations-of-prediction},
which were themselves independent.  (This area of the literature is more than
usually tangled.)

Efficient {\bold reconstruction algorithms} or {\bold discovery procedures}
exist for building CSMs \cite{CSSR-for-UAI} and OOMs
\cite{Jaeger-operator-models} directly from data.  (There is currently no such
discovery procedure for PSRs, though there are parameter-estimation algorithms
\cite{learning-PSRs}.)  These algorithms are reliable, in the sense that, given
enough data, the probability that they build the wrong set of states becomes
arbitrarily small.  Experimentally, selecting an HMM architecture through
cross-validation never does better than reconstruction, and often much worse
\cite{CSSR-for-UAI}.

While these models are more powerful than VLMMs, there are still many
stochastic processes which cannot be represented in this form; or, rather,
their representation requires an infinite number of states
\cite{Upper-thesis,Dupont-et-al-automata-and-HMMs}.  This is mathematically
unproblematic, though reconstruction will then become much harder.  (For
technical reasons, it seems likely to be easier to carry through for OOMs or
PSRs than for CSMs.)  In fact, one can show that these techniques would work
straight-forwardly on continuous-valued, continuous-time processes, if only we
knew the necessary conditional distributions
\cite{Knight-predictive-view,Jaeger-characterizing-distributions}.  Devising a
reconstruction algorithm suitable for this setting is an extremely challenging
and completely unsolved problem; even parameter estimation is difficult, and
currently only possible under quite restrictive assumptions
\cite{Jaeger-learning-continuous-valued}.

\subsubsection{Generating Partitions}
\label{sec:generating-partitions}

So far, everything has assumed that we are either observing truly discrete
quantities, or that we have a fixed discretization of our continuous
observations.  In the latter case, it is natural to wonder how much difference
the discretization makes.  The answer, it turns out, is {\em quite a lot};
changing the partition can lead to completely different symbolic dynamics
\cite{JPC-unreconstructable,Bollt-et-al-validity,%
  Bollt-et-al-misplaced-partition}.  How then might we choose a {\em good}
partition?

Nonlinear dynamics provides an answer, at least for deterministic systems, in
the idea of a {\bold generating partition} \cite{Badii-Politi,Kitchens}.
Suppose we have a continuous state $x$ and a deterministic map on the state
$F$, as in \S \ref{sec:state-space-picture}.  Under a partitioning $\phi$, each
point $x$ in the state space will generate an infinite sequence of symbols,
$\Phi(x)$, as follows: $\phi(x), \phi(F(x)), \phi(F^2(x)), \ldots$.  The
partition $\phi$ is generating if each point $x$ corresponds to a {\em unique}
symbol sequence, i.e., if $\Phi$ is invertible.  Thus, no information is lost
in going from the continuous state to the discrete symbol sequence\footnote{An
  alternate definition \cite{Badii-Politi} looks at the entropy rate (\S
  \ref{sec:info-theory}) of the symbol sequences: a generating partition is one
  which maximizes the entropy rate, which is the same as maximizing the extra
  information about the initial condition $x$ provided by each symbol of the
  sequence $\Phi(x)$.}.  While one must know the continuous map $F$ to
determine exact generating partitions, there are reasonable algorithms for
approximating them from data, particularly in combination with embedding
methods
\cite{Fraser-Swinney-independent-coords,Kennel-Buhl-estimating-partitions,%
  Hirata-et-al-estimating-partition}.  When the underlying dynamics are
stochastic, however, the situation is much more complicated
\cite{JPC-Packard-noisy-chaos}.

\section{Cellular Automata}
\label{sec:cas}

{\bold Cellular automata} are one of the more popular and distinctive classes
of models of complex systems.  Originally introduced by von Neumann as a way of
studying the possibility of mechanical self-reproduction, they have established
niches for themselves in foundational questions relating physics to computation
in statistical mechanics, fluid dynamics, and pattern formation.  Within that
last, perhaps the most relevant to the present purpose, they have been
extensively and successfully applied to physical and chemical pattern
formation, and, somewhat more speculatively, to biological development and to
ecological dynamics.  Interesting attempts to apply them to questions like the
development of cities and regional economies lie outside the scope of this
chapter.

\subsection{A Basic Explanation of CA}

Take a board, and divide it up into squares, like a chess-board or
checker-board.  These are the cells.  Each cell has one of a finite number of
distinct colors --- red and black, say, or (to be patriotic) red, white and
blue.  (We do not allow continuous shading, and every cell has just one color.)
Now we come to the ``automaton'' part.  Sitting somewhere to one side of the
board is a clock, and every time the clock ticks the colors of the cells
change.  Each cell looks at the colors of the nearby cells, and its own color,
and then applies a definite rule, the {\bold transition rule}, specified in
advance, to decide its color in the next clock-tick; and all the cells change
at the same time.  (The rule can say ``Stay the same.'')  Each cell is a sort
of very stupid computer --- in the jargon, a {\bold finite-state automaton} ---
and so the whole board is called a {\bold cellular automaton}, or CA.  To run
it, you color the cells in your favorite pattern, start the clock, and stand
back.

Let us follow this concrete picture with one more technical and abstract.  The
cells do not have to be colored, of course; all that's important is that each
cell is in one of a finite number of states at any given time.  By custom
they're written as the integers, starting from 0, but any ``finite alphabet''
will do.  Usually the number of states is small, under ten, but in principle
any finite number is allowed.  What counts as the ``nearby cells'', the {\bold
  neighborhood}, varies from automaton to automaton; sometimes just the four
cells on the principle directions, sometimes the corner cells, sometimes a
block or diamond of larger size; in principle any arbitrary shape.  You do not
need to stick to a chess-board; you can use any regular pattern of cells which
will fill the plane (or ``tessellate'' it; an old name for cellular automata is
{\bold tessellation structures}).  And you do not have to stick to the plane;
any number of dimensions is allowed.  There are various tricks for handling the
edges of the space; the one which has ``all the advantages of theft over honest
toil'' is to assume an infinite board.

\paragraph{Cellular Automata as Parallel Computers}

CA are synchronous massively parallel computers, with each cell being a finite
state transducer, taking input from its neighbors and making its own state
available as output.  From this perspective, the remarkable thing about CA is
that they are computationally universal, able to calculate any (classically)
computable function; one can use finite-state machines, the least powerful kind
of computer, to build devices equivalent to Turing machines, the most powerful
kind of computer.  The computational power of different physically-motivated
CA is an important topic in complex systems
\cite{Moore-majority-vote,Moore-Nordahl-lattice-gases}, though it must be
confessed that CA with very different computational powers can have very
similar behavior in most respects.

\paragraph{Cellular Automata as Discrete Field Theories}

From the perspective of physics, a CA is a ``digitized'' classical field
theory, in which space, time and the field (state) are all discrete.  Thus
fluid mechanics, continuum mechanics, and electromagnetism can all be simulated
by CA\footnote{Quantum versions of CA are an active topic of investigation,
  but unlikely to be of biological relevance \cite{Ilachinski-discrete}.}
typically, however, the physical relevance of a CA comes not from accurately
simulating some field theory at the microscopic level, but from the large-scale
phenomena they generate.

\begin{figure}
\begin{center}
\resizebox{2in}{!}{\includegraphics{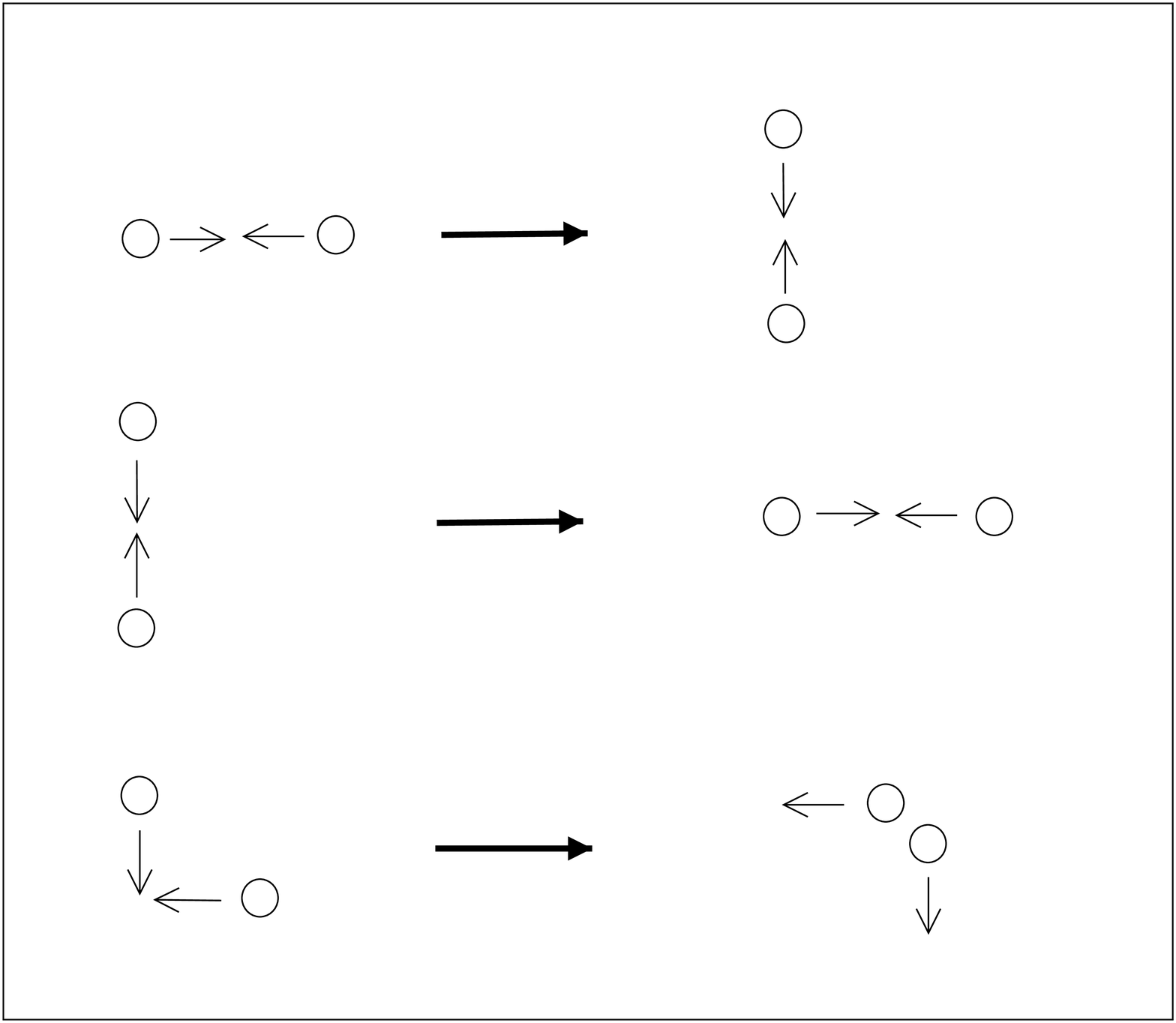}}
\end{center}
\caption{Collisions in the HPP lattice gas rule.  Horizontal collisions produce
  vertically-moving particles (top) and vice versa (middle).  Particles moving
  at right angles pass by each other unchanged (bottom, omitting the
  reflections and rotations of this figure).}
\label{fig:lattice-gas}
\end{figure}

Take, for example, simulating fluid mechanics, where CA are also called {\bold
  lattice gases} or {\bold lattice fluids}.  In the ``HPP'' \cite{HPP} rule, a
typical lattice gas with a square grid, there are four species of ``fluid
particle'', which travel along the four principal directions.  If two cells
moving in opposite directions try to occupy the same location at the same time,
they collide, and move off at right angles to their original axis (Figure
\ref{fig:lattice-gas}).  Each cell thus contains only an integer number of
particles, and only a discrete number of values of momentum are possible.  If
one takes averages over reasonably large regions, however, then density and
momentum approximately obey the equations of continuous fluid mechanics.
Numerical experiments show that this rule reproduces many fluid phenomena, such
as diffusion, sound, shock-waves, etc.  However, with this rule, the agreement
with fluid mechanics is only approximate.  In particular, the square lattice
makes the large-scale dynamics anisotropic, which is unphysical.  This in turn
can be overcome in several ways --- for instance, by using a hexagonal lattice
\cite{FHP}.  The principle here --- get the key parts of the small-scale
``microphysics'' right, and the interesting ``macrophysics'' will take care of
itself --- is extensively applied in studying pattern formation, including such
biologically-relevant phenomena as phase separation
\cite{Rothman-Zaleski-text}, excitable media
\cite{Fisch-Gravner-Griffeath-threshold-range}, and the self-assembly of
micelles \cite{Nilsson-et-al-constructive-molecular-dynamics,%
  Nilsson-Rasmussen-CA-for-self-assembly}.

\section{Agent-Based Models}
\label{sec:abms}

If there is any {\em one} technique associated with complex systems science, it
is agent-based modeling.  An agent-based model is a computational model which
represents individual agents and their collective behavior.  What, exactly, do
we mean by ``agent''?  Stuart Kauffman has offered\footnote{In a talk at the
  Santa Fe Institute, summer of 2000; the formula does not seem to have been
  published.} the following apt definition: ``An agent is a thing which does
things to things''.  That is, an agent is a persistent thing which has some
{\em state} we find worth representing, and which interacts with other agents,
mutually modifying each others' states.  The components of an agent-based model
are a collection of agents and their states, the rules governing the
interactions of the agents, and the environment within which they live.  (The
environment need not be represented in the model if its effects are constant.)
The state of an agent can be arbitrarily simple, say just position, or the
color of a cell in a CA.  (At this end, agent-based models blend with
traditional stochastic models.)  States can also be extremely complicated,
including, possibly, sophisticated internal models of the agent's world.

Here is an example to make this concrete.  In epidemiology, there is a classic
kind of model of the spread of a disease through a population called an ``SIR''
model \cite[\S 4.4]{Bartlett-stochastic}.  It has three classes of people ---
the susceptible, who have yet to be exposed to the disease; the infected, who
have it and can pass it on; and the resistant or recovered, who have survived
the disease and cannot be re-infected.  A traditional approach to an SIR model
would have three variables, namely the number of people in each of the three
categories, $S(t), I(t), R(t)$, and would have some deterministic or stochastic
dynamics in terms of those variables.  For instance, in a deterministic SIR
model, one might have
\begin{eqnarray}
\label{eqn:SIR-S}
S(t+1) - S(t) & = &  -a\left(\frac{I(t)}{S(t)+I(t)+R(t)}\right)S(t)\\
\label{eqn:SIR-I}
I(t+1) - I(t) & = &  \left[a \frac{S(t)}{S(t)+I(t)+R(t)} - b - c\right]I(t)\\
\label{eqn:SIR-R}
R(t+1) - R(t) & = & bI(t)
\end{eqnarray}
which we could interpret by saying that (i) the probability of a susceptible
person being infected is proportional to the fraction of the population which
is already infected, (ii) infected people get better at a rate $b$ and (iii)
infected people die at a rate $c$.  (This is not a particularly {\em realistic}
SIR model.)  In a stochastic model, we would treat the right hand sides of
(\ref{eqn:SIR-S})--(\ref{eqn:SIR-R}) as the mean changes in the three
variables, with (say) Poisson-distributed fluctuations, taking
care that, e.g., the fluctuation in the $a\frac{I}{R+S+I}$ term in
(\ref{eqn:SIR-S}) is the same as that in (\ref{eqn:SIR-I}).  The thing to note
is that, whether deterministic or stochastic, the whole model is cast in terms
of the aggregate quantities $S$, $I$ and $R$, and those aggregate variables are
what we would represent computationally.

In an agent-based model of the same dynamics, we would represent {\em each}
individual in the population as a distinct agent, which could be in one of
three states, S, I and R.  A simple interaction rule would be that at each
time-step, each agent selects another from the population entirely at random.
If a susceptible agent (i.e., one in state S) picks an infectious agent (i.e.,
one in state I), it becomes infected with probability $a$.  Infectious agents
die with probability $b$ and recover with probability $c$; recovered agents
never change their state.

So far, we have merely reproduced the stochastic version of
(\ref{eqn:SIR-S})--(\ref{eqn:SIR-R}), while using many more variables.  The
power of agent-based modeling only reveals itself when we implement more
interesting interaction rules.  For instance, it would be easy to assign each
agent a position, and make two agents more likely to interact if they are
close.  We could add visible symptoms which are imperfectly associated with the
disease, and a tendency not to interact with symptomatic individuals.  We could
make the degree of aversion to symptomatic agents part of the agents' state.
All of this is easy to implement in the model, even in combination, but {\em
  not} easy to do in a more traditional, aggregated model.  Sometimes it would
be all but impossible; an excellent case in point is the highly sophisticated
model of HIV epidemiology produced by Jacquez, Koopman, Simon and collaborators
\cite{Jacquez-et-al-on-HIV,Koopman-et-al-on-HIV}, incorporating multiple routes
of transmission, highly non-random mixing of types, and time-varying
infectiousness.

Agent-based models steer you towards representing individuals, their behaviors
and their interactions, rather than aggregates and their dynamics.  Whether
this is a good thing depends, of course, on what you know, and what you hope to
learn.  If you know a lot about individuals, agent-based models can help you
leverage that knowledge into information about collective dynamics.  This is
particularly helpful if the population is heterogeneous, since you can
represent the different types of individuals in the population by different
states for agents.  This requires a bit of effort on your part, but often not
nearly so much as it would to represent the heterogeneity in an aggregated
model.  Conversely, if you think you have the collective dynamics down, an ABM
will let you check whether a candidate for an individual-level mechanism really
will produce them.  (But see \S \ref{sec:evaluation}, below.)

Ideally, there are no ``mass nouns'' in an ABM, nothing represented by a
smeared-out ``how much'': everything should be represented by some definite
number of distinctly-located agents.  At most, some aggregate variables may be
stuffed into the environment part of the model, but only simple and homogeneous
ones.  Of course, the {\em level} of disaggregation at which it is useful to
call something an agent is a matter for particular applications, and need not
be the same for every agent in a model.  (E.g., one might want to model an
entire organ as a single agent, while another, more interesting organ is broken
up into multiple interacting agents, along anatomical or functional lines.)
Sometimes it's just not practical to represent everything which we know is an
individual thing by its own agent: imagine trying to do chemical thermodynamics
by tracking the interactions of a mole of molecules.  Such cases demand either
giving up on agent-based modeling (fortunately, the law of mass action works
pretty well in chemistry), or using fictitious agents that represent
substantial, but not too large, collections of individuals.

Models describing the collective dynamics of aggregate variables are sometimes
called ``equation-based models'', in contrast to agent-based models.  This is
sloppy, however: it is always possible, though generally tedious and
unilluminating, to write down a set of equations which describe the dynamics of
an agent-based model.  Rather than drawing a false contrast between agents and
equations, it would be better to compare ABMs to ``aggregate models'',
``collective models'' or perhaps ``factor models''.

\subsection{Computational Implementation: Agents are Objects}

The nicest way to computationally implement the commitment of distinctly
representing each agent, is to make agents {\bold objects}, which are, to
over-simplify slightly, data structures which have internal states, and
interact with each other by passing messages.  While objects are not necessary
for agent-based models, they do make programming them {\em much} easier,
especially if the agents have much more state than, say, just a position and a
type.  If you try to implement models with sophisticated agents without using
objects, the odds are good that you will find yourself re-inventing well-known
features of object-oriented programming.  (Historically, object-oriented
programming {\em began} with languages for simulation modeling
\cite{Budd-understanding-oop}.)  You might as well save your time, and do those
things {\em right}, by using objects in the first place.

Generally speaking, computational implementations of ABMs contain many
non-agent objects, engaged in various housekeeping tasks, or implementing the
functions agents are supposed to perform.  For instance, an agent, say a rat,
might be supposed to memorize a sequence, say of turns in a maze.  One way of
implementing this would be to use a linked list, which is an object itself.
Such objects do not represent actual features of the {\em model}, and it should
be possible to vary them without interfering with the model's behavior.  Which
objects are picked out as agents is to some degree a matter of convenience and
taste.  It is common, for instance, to have mobile agents interacting on a
static environment.  If the environment is an object, modelers may or may not
speak of it as an ``environment agent,'' and little seems to hinge on whether
or not they do.

There are several programming environments designed to facilitate agent-based
modeling.  Perhaps the best known of these is \textsc{Swarm}
(\url{www.swarm.org}), which works very flexibly with several languages, is
extensively documented, and has a large user community, though it presently
(2004) lacks an institutional home.  \textsc{RePast}, while conceptually
similar, is open-source (\url{repast.sourceforge.net}) and is associated with
the University of Chicago.  \textsc{StarLogo}, and it successor,
\textsc{NetLogo} (\url{ccl.sesp.northwestern.edu/netlogo}), are extensions of
the popular \textsc{Logo} language to handle multiple interacting ``turtles'',
i.e., agents.  Like Logo, children can learn to use them
\cite{Resnick-turtles}, but they are fairly easy for adults, too, and certainly
give a feel for working with ABMs.

\subsection{Three Things Which Are Not Agent-Based Models}

Not everything which involves the word ``agent'' is connected to agent-based
modeling.

{\bold Representative agent models} are not ABMs.  In these models, the
response of a population to environment conditions is found by picking out a
{\em single} typical or representative agent, determining their behavior, and
assuming that everyone else does likewise.  This is sometimes reasonable, but
it's clearly diametrically opposed to what an ABM is supposed to be.

{\bold Software agents} are not ABMs.  Software agents are a very useful and
rapidly developing technology \cite[chapter 2]{Brown-Duguid-SLoI}; an agent,
here, is roughly a piece of code which interacts with other software and with
pieces of the real world autonomously.  Agents index the Web for search
engines, engage in automated trading, and help manage parts of the North
American electrical power grid, among other things.  Some agent software
systems are {\em inspired} by ABMs
\cite{Bonabeau-Dorigo-Theraulaz-swarm-intelligence}.  When one wants to model
their behavior, an ABM is a natural tool (but not the only one by any means:
see \cite{Lerman-design-and-analysis}).  But a set of software agents running
the Michigan power grid is not a {\em model} of anything, it's {\em doing}
something.

Finally, {\bold multi-agent systems} \cite{Ossowski-artificial-agent-societies}
and {\bold rational agents} \cite{Wooldridge-reasoning} in artificial
intelligence are not ABMs.  The interest of this work is in understanding, and
especially {\em designing}, systems capable of sophisticated, autonomous
cognitive behavior; many people in this field would restrict the word ``agent''
to apply only to things capable, in some sense, of having ``beliefs, desires
and intentions''.  While these are certainly complex systems, they are not
usually intended to be {\em models} of anything else.  One can, of course,
press them into service as models
\cite{Jonker-et-al-intentions-in-biochemistry}, but generally this will be no
more than a heuristic device.

\subsection{The Simplicity of Complex Systems Models}

One striking feature of agent-based models, and indeed of complex systems
models in general, is how {\em simple} they are.  Often, agents have only a few
possible states, and only a handful of kinds of interaction.  This practice has
three motivations.  (i) A model as detailed as the system being studied would
be as hard to understand as that system.  (ii) Many people
working in complex systems science want to show that a certain set of
mechanisms are sufficient to generate some phenomenon, like cooperation among
unrelated organisms, or the formation of striped patterns.  Hence using simple
models, which contain only those mechanisms, makes the case.  (iii) Statistical
physicists, in particular, have a long tradition of using highly simplified
models as caricatures of real systems.

All three motives are appropriate, in their place.  (i) is completely
unexceptionable; abstracting away from irrelevant detail is always worthwhile,
so long as it really is irrelevant.  (ii) is also fair enough, though one
should be careful that the mechanisms in one's model can still generate the
phenomenon when they interact with {\em other} effects as well.  (iii) works
very nicely in statistical physics itself, where there are powerful
mathematical results relating to the renormalization group
\cite{Chaikin-Lubensky} and bifurcation theory \cite{Cross-Hohenberg} which
allow one to extract certain kinds of {\em quantitative} results from
simplified models which share certain {\em qualitative} characteristics with
real systems.  (We have seen a related principle when discussing cellular
automata models above.)  There is, however, little reason to think that these
universality results apply to most complex systems, let alone ones with
adaptive agents!

\section{Evaluating Models of Complex Systems}
\label{sec:evaluation}

We do not build models for their own sake; we want to see what they do, and we
want to compare what they do both to reality and to other models.  This kind of
evaluation of models is a problem for all areas of science, and as such little
useful general advice can be given.  However, there are some issues which are
peculiar to models of complex systems, or especially acute for them, and I will
try to provide some guidance here, moving from figuring out just what your
model does, to comparing your model to data, to comparing it to other models.

\subsection{Simulation}

The most basic way to see what your model does is to run it; to do a
simulation.  Even though a model is entirely a human construct, every aspect of
its behavior following logically from its premises and initial conditions, the
frailty of human nature is such that we generally cannot perceive those
consequences, not with any accuracy.  If the model involves a large number of
components which interact strongly with each other --- if, that is to say, it's
a good model of a complex system --- our powers of deduction are generally
overwhelmed by the mass of relevant, interconnected detail.  Computer
simulation then comes to our aid, because computers have no trouble remembering
large quantities of detail, nor in following instructions.

\subsubsection{Direct Simulation}

Direct simulation --- simply starting the model and letting it go --- has two
main uses.  One is to get a sense of the typical behavior, or of the range of
behavior.  The other, more quantitative, use is to determine the distribution
of important quantities, including time series.  If one randomizes initial
conditions, and collects data over multiple runs, one can estimate the
distribution of desired quantities with great accuracy.  This is exploited in
the time series method of surrogate data (above), but the idea applies quite
generally.

Individual simulation runs for models of complex systems can be reasonably
expensive in terms of time and computing power; large numbers of runs, which
are really needed to have confidence in the results, are correspondingly more
costly.  Few things are more dispiriting than to expend such quantities of time
and care, only to end up with ambiguous results.  It is almost always
worthwhile, therefore, to carefully think through what you want to measure, and
why, before running anything.  In particular, if you are trying to judge the
merits of competing models, effort put into figuring out how and where they are
{\em most} different will generally be well-rewarded.  The theory of
experimental design offers extensive guidance on how to devise informative
series of experiments, both for model comparison and for other purposes, and by
and large the principles apply to simulations as well as to real experiments.

\subsubsection{Monte Carlo Methods}

{\bold Monte Carlo} is the name of a broad, slightly indistinct family for
using random processes to estimate deterministic quantities, especially the
properties of probability distributions.  A classic example will serve to
illustrate the basic idea, on which there are many, many refinements.

Consider the problem of determining the area $A$ under an curve given by a
known but irregular function $f(x)$.  In principle, you could integrate $f$ to
find this area, but suppose that numerical integration is infeasible for some
reason.  (We will come back to this point presently.)  A Monte Carlo solution
to this problem is as follows: pick points at random, uniformly over the
square.  The probability $p$ that a point falls in the shaded region is equal
to the fraction of the square occupied by the shading: $p = A/s^2$.  If we pick
$n$ points independently, and $x$ of them fall in the shaded region, then $x/n
\rightarrow p$ (by the law of large numbers), and $s^2 x/n \rightarrow A$.
$s^2 x/n$ provides us with a stochastic estimate of the integral.  Moreover,
this is a probably approximately correct (\S \ref{sec:PAC-and-SRM}) estimate,
and we can expect, from basic probability theory, that the standard deviation
of the estimate around its true value will be proportional to $n^{-1/2}$, which
is not bad\footnote{A simple argument just invokes the central limit theorem.
  The number of points falling within the shaded region has a binomial
  distribution, with success parameter $p$, so asymptotically $x/n$ has a
  Gaussian distribution with mean $p$ and standard deviation $\sqrt{p(1-p)/n}$.
  A non-asymptotic result comes from Chernoff's inequality
  \cite{Vidyasagar-learning}, which tells us that, for all $n$, $\Prob(|x/n -
  p| \geq \epsilon) < 2e^{-2n{\epsilon}^2}$.}.  However, when faced with such a
claim, one should always ask what the proportionality constant is, and whether
it is the best achievable.  Here it is not: the equally simple, if less visual,
scheme of just picking values of $x$ uniformly and averaging the resulting
values of $f(x)$ always has a smaller standard deviation \cite[chapter
  5]{Hammersley-Handscomb}.

This example, while time-honored and visually clear, does not show Monte Carlo
to its best advantage; there are few one-dimensional integrals which cannot be
done better by ordinary, non-stochastic numerical methods.  But numerical
integration becomes computationally intractable when the domain of integration
has a large number of dimensions, where ``large'' begins somewhere between four
and ten.  Monte Carlo is much more indifferent to the dimensionality of the
space: we could replicate our example with a 999-dimensional hyper-surface in a
1000-dimensional space, and we'd still get estimates that converged like
$n^{-1/2}$, so achieving an accuracy of $\pm \epsilon$ will require evaluating
the function $f$ only $O(\epsilon^{-2})$ times.

Our example was artificially simple in another way, in that we used a uniform
distribution over the entire space.  Often, what we want is to compute the
expectation of some function $f(x)$ with a non-uniform probability $p(x)$.
This is just an integral, $\int{f(x)p(x) dx}$, so we could sample points
uniformly and compute $f(x)p(x)$ for each one.  But if some points have very
low probability, so they only make a small contribution to the integral,
spending time evaluating the function there is a bit of a waste.  A better
strategy would be to pick points according to the actual probability
distribution.  This can sometimes be done directly, especially if $p(x)$ is of
a particularly nice form.  A very general and clever indirect scheme is as
follows \cite{Metropolis-et-al-Monte-Carlo}.  We want a whole sequence of
points, $x_1, x_2, \ldots x_n$.  We pick the first one however we like, and
after that we pick successive points according to some Markov chain: that is,
the distribution of $x_{i+1}$ depends only on $x_i$, according to some fixed
function $q(x_i, x_{i+1})$.  Under some mild conditions\footnote{The chain
  needs to be irreducible, meaning one can go from any point to any other
  point, and positive recurrent, meaning that there's a positive probability of
  returning to any point infinitely often.}, the distribution of $x_t$
approaches a stationary distribution $q^{*}(x)$ at large times $t$.  If we
could ensure that $q^{*}(x) = p(x)$, we know that the Markov chain was
converging to our distribution, and then, by the ergodic theorem, averaging
$f(x)$ along a trajectory would give the expected value of $f(x)$.  One way to
ensure this is to use the ``detailed balance'' condition of the invariant
distribution, that the total probability of going from $x$ to $y$ must equal
the total probability of going the other way:
\begin{eqnarray}
p(x) q(x,y) & = & p(y) q(y,x)\\
\label{eqn:detailed-condition}
\frac{q(x,y)}{q(y,x)} & = & \frac{p(y)}{p(x)} \equiv h(x,y)
\end{eqnarray}
So now we just need to make sure that (\ref{eqn:detailed-condition}) is
satisfied.  One way to do this is to set $q(x,y) =
\min{\left(1,h(x,y)\right)}$; this was the original proposal of Metropolis {\em
  et al}.  Another is $q(x,y) = \frac{h(x,y)}{1+h(x,y)}$.  This method is what
physicists usually mean by ``Monte Carlo'', but statisticians call it {\bold
  Markov chain Monte Carlo}, or ``MCMC''.  While we can now estimate the
properties of basically arbitrary distributions, we no longer have independent
samples, so evaluating the accuracy of our estimates is no longer a matter of
{\em trivial} probability\footnote{Unless our choices for the transition
  probabilities are fairly perverse, the central limit theorem still holds, so
  asymptotically our estimate still has a Gaussian distribution around the true
  value, and still converges as $N^{-1/2}$ for large enough $N$, but
  determining what's ``large enough'' is trickier.}.  An immense range of
refinements have been developed over the last fifty years, addressing these and
other points; see the further reading section for details.

Keep in mind that Monte Carlo is a stochastic simulation method only in a
special sense --- it simulates the probability distribution $p(x)$, {\em not}
the mechanism which generated that distribution.  The dynamics of Markov chain
Monte Carlo, in particular, often bear no resemblance whatsoever to those of
the real system\footnote{An important exception is the case of equilibrium
  statistical mechanics, where the dynamics under the Metropolis algorithm {\em
    are} like the real dynamics.}.  Since the point of Monte Carlo is to tell
us about the properties of $p(x)$ (what is the expectation value of this
function? what is the probability of configurations with this property? etc.),
the actual trajectory of the Markov chain is of no interest.  This point
sometimes confuses those more used to direct simulation methods.

\subsection{Analytical Techniques}

Naturally enough, analytical techniques are not among the tools which first
come to mind for dealing with complex systems; in fact, they often do not come
to mind at all.  This is unfortunate, because a lot of intelligence has been
devoted to devising approximate analytical techniques for classes of models
which include many of those commonly used for complex systems.  A general
advantage of analytical techniques is that they are often fairly insensitive to
many details of the model.  Since any model we construct of a complex system is
almost certainly much simpler than the system itself, a great many of its
details are just wrong.  If we can extract non-trivial results which are
insensitive to those details, we have less reason to worry about this.

One particularly useful, yet neglected, body of approximate analytical
techniques relies on the fact that many complex systems models are Markovian.
In an agent-based model, for instance, the next state of an agent generally
depends only on its present state, and the present states of the agents it
interacts with.  If there is a fixed interaction graph, the agents form a
Markov random field on that graph.  There are now very powerful and
computationally efficient methods for evaluating many properties of Markov
chains \cite{Honerkamp-stochastic,Bremaud-markov-chains}, Markov random fields
\cite{Beckerman-adaptive}, and (closely related) graphical models
\cite{Jordan-learning-in-graphical-models} {\em without} simulation.  The
recent books of Peyton Young \cite{Young-strategy-structure} and Sutton
\cite{Sutton-technology-and-market-structure} provide nice instances of using
analytical results about Markov processes to solve models of complex social
systems, without impractical numerical experiments.

\subsection{Comparisons with Data}

\subsubsection{General issues}

We can only compare particular aspects of a model of a system to particular
kinds of data about that system.  The most any experimental test can tell us,
therefore, is how similar the model is to the system {\em in that respect}.
One may think of an experimental comparison as a test for a {\em particular}
kind of {\em error}, one of the infinite number of mistakes which we could make
in building a model.  A good test is one which is very likely to alert us to an
error, if we have made it, but not otherwise \cite{Mayo-error}.

These ought to be things every school-child knows about testing hypotheses.  It
is very easy, however, to blithely ignore these truisms when confronted with,
on the one hand, a system with many strongly interdependent parts, and, on the
other hand, a model which tries to mirror that complexity.  We must decide
which features of the model {\em ought} to be similar to the system, and how
similar.  It is important not only that our model be able to adequately
reproduce those phenomena, but that it not entail badly distorted or
non-existent phenomena in other respects.

\subsubsection{Two Stories and Some Morals}

Let me give two examples from very early in the study of complex systems, which
nicely illustrate some fundamental points.

The first has to do with pattern formation in chemical oscillators
\cite{Epstein-Pojman}.  Certain mixtures of chemicals in aqueous solution,
mostly famously the Belusov-Zhabotinsky reagent, can not only undergo cyclic
chemical reactions, but will form rotating spiral waves, starting from an
initial featureless state.  This is a visually compelling example of
self-organization, and much effort has been devoted to understanding it.  One
of the more popular early models was the ``Brusselator'' advanced by Prigogine
and his colleagues at the Free University of Brussels; many similarly-named
variants developed.  Brusselator-type models correctly predicted that these
media would support spiral waves.  They all, further, predicted that the
spirals would form only when the homogeneous configuration was unstable, and
that then they would form spontaneously.  It proved very easy, however, to
prepare the Belusov-Zhabotisnky reagent in such a way that it was ``perfectly
stable in its uniform quiescence'', yet still able to produce spiral waves if
excited (e.g., by being touched with a hot wire)
\cite{Winfree-time-breaks-down}. The Brusselator and its variants were simply
unable to accommodate these phenomena, and had to be discarded in favor of
other models.  The fact that these were qualitative results, rather than
quantitative ones, if anything made it more imperative to get rid of the
Brusselator.

The second story concerns the work of Varela and Maturana on ``autopoesis''.
In a famous paper \cite{Varela-Maturana-Uribe}, they claimed to exhibit a
computational model of a simple artificial chemistry where membranes not only
formed spontaneously, but a kind of metabolism self-organized to sustain the
membranes.  This work influenced not just complex systems science but
theoretical biology, psychology, and even sociology
\cite{Luhmann-social-systems}.  When, in the 1990s, McMullin made the first
serious effort to reproduce the results, based on the description of the model
in the paper, that description proved {\em not} to match the published
simulation results.  The discrepancy was only resolved by the fortuitous
rediscovery of a mass of papers, including Fortran code, that Varela had left
behind in Chile when forced into exile by the fascist regime.  These revealed a
crucial change in one particular reaction made all the difference between
successful autopoesis and its absence.  (For the full story, see
\cite{McMullin-independent,McMullin-Varela}.)  Many similar stories could be
told of other models in complex systems \cite{Revisiting-edge-of-chaos}; this
one is distinguished by McMullin's unusual tenacity in trying to replicate the
results, Varela's admirable willingness to assist him, and the happy ending.

The story of autopoesis is especially rich in morals.  (1) Replication is
essential.  (2) It is a good idea to share not just data but programs.  (3)
{\em Always} test the robustness of your model to changes in its parameters.
(This is fairly common.)  (4) {\em Always} test your model for robustness to
small changes in qualitative assumptions.  If your model calls for a given
effect, there are usually several mechanisms which could accomplish it.  If it
does not matter which mechanism you actually use, the result is that much more
robust.  Conversely, if it does matter, the over-all adequacy of the model can
be tested by checking whether {\em that} mechanism is actually present in the
system.  Altogether too few people perform such tests.

\subsubsection{Comparing Macro-data and Micro-models}

Data are often available only about large aggregates, while models, especially
agent-based models, are about individual behavior.  One way of comparing such
models to data is to compute the necessary aggregates, from direct simulation,
Monte Carlo, etc.  The problem is that many different models can give the same
aggregated behavior, so this does not provide a powerful test between different
models.  Ideally, we'd work back from aggregate data to individual behaviors,
which is known, somewhat confusingly, as {\bold ecological inference}.  In
general, the ecological inference problem itself does not have a unique
solution.  But the aggregate data, if used intelligently, can often put fairly
tight constraints on the individual behaviors, and micro-scale can be directly
checked against those constraints.  Much of the work here has been done by
social scientists, especially American political scientists concerned with
issues arising from the Voting Rights Act
\cite{Schuessler-ecological-inference}, but the methods they have developed are
very general, and could profitably be applied to agent-based models in the
biological sciences, though, to my knowledge, they have yet to be.

\subsection{Comparison to Other Models}

Are there other ways of generating the data?  There generally are, at least if
``the data'' are some very gross, highly-summarized pattern.  This makes it
important to look for differential signatures, places where discrepancies
between different generative mechanisms give one some {\em leverage}.  Given
two mechanisms which can both account for our phenomenon, we should look for
some {\em other} quantity whose behavior will be different under the two
hypotheses.  Ideally, in fact, we would look for the statistic on which the two
kinds of model are {\em most} divergent.  The literature on experimental design
is relevant here again, since it considers such problems under the heading of
{\bold model discrimination}, seeking to maximize the power of experiments (or
simulations) to distinguish between different classes of models
\cite{Atkinson-Donev,Borowiak-model-discrimination}.

Perhaps no aspect of methodology is more neglected in complex systems science
than this one.  While it is always perfectly legitimate to announce a new
mechanism as {\em a} way of generating a phenomenon, it is far too common for
it to be called {\em the} way to do it, and vanishingly rare to find an
examination of how it {\em differs} from previously-proposed mechanisms.
Newman and Palmer's work on extinction models \cite{MEJN-Palmer-extinctions}
stands out in this regard for its painstaking examination of the ways of
discriminating between the various proposals in the literature.

\section{Information Theory}
\label{sec:info-theory}

Information theory began as a branch of communications engineering, quantifying
the length of codes needed to represent randomly-varying signals, and the rate
at which data can be transmitted over noisy channels.  The concepts needed to
solve these problems turn out to be quite fundamental measures of the
uncertainty, variability, and the interdependence of different variables.
Information theory thus is an important tool for studying complex systems, and
in addition is indispensable for understanding complexity measures (\S
\ref{sec:complexity}).

\subsection{Basic Definitions}

Our notation and terminology follows that of Cover and Thomas's standard
textbook \cite{Cover-and-Thomas}.

Given a random variable $X$ taking values in a discrete set $\mathcal{A}$, the
{\bold entropy} or {\bold information content} $H[X]$ of $X$ is
\begin{eqnarray}
H[X] & \equiv & -\sum_{a\in\mathcal{A}}{\Prob(X=a)\log_2{\Prob(X=a)}} ~.
\end{eqnarray}
$H[X]$ is the expectation value of $-\log_2{\Prob(X)}$.  It represents the
uncertainty in $X$, interpreted as the mean number of binary distinctions
(bits) needed to identify the value of $X$.  Alternately, it is the minimum
number of bits needed to encode or describe $X$.  Note that $H[X] = 0$ if and
only if $X$ is (almost surely) constant.

The {\bold joint entropy} $H[X,Y]$ of two variables $X$ and $Y$ is the entropy
of their joint distribution:
\begin{eqnarray}
H[X,Y] & \equiv & -\sum_{a\in\mathcal{A},b\in\mathcal{B}}{\Prob(X=a,Y=b)\log_2{\Prob(X=a,Y=b)}} ~.
\end{eqnarray}

The {\bold conditional entropy} of $X$ given $Y$ is
\begin{eqnarray}
H[X|Y] & \equiv & H[X,Y] - H[Y] ~.
\end{eqnarray}
$H[X|Y]$ is the average uncertainty remaining in $X$, given a knowledge of $Y$.

The {\bold mutual information} $I[X;Y]$ between $X$ and $Y$ is
\begin{eqnarray}
I[X;Y] & \equiv &  H[X] - H[X|Y] ~.
\end{eqnarray}
It gives the reduction in $X$'s uncertainty due to knowledge of $Y$ and is
symmetric in $X$ and $Y$.  We can also define higher-order mutual informations,
such as the third-order information $I[X;Y;Z]$,
\begin{eqnarray}
I[X;Y;Z] & \equiv & H[X] + H[Y] + H[Z] - H[X,Y,Z]
\end{eqnarray}
and so on for higher orders.  These functions reflect the joint dependence
among the variables.

Mutual information is a special case of the {\bold relative entropy}, also
called the {\bold Kullback-Leibler divergence} (or {\bold distance}).  Given
two {\em distributions} (not variables), $\Dist$ and $\AltDist$, the entropy of
$\AltDist$ relative to $\Dist$ is
\begin{eqnarray}
D(\Dist\|\AltDist) &\equiv & \sum_{x}{\Dist(x)\log{\frac{\Dist(x)}{\AltDist(x)}}}
\end{eqnarray}
$D$ measures how far apart the two distributions are, since $D(\Dist||\AltDist)
\geq 0$, and $D(\Dist\|\AltDist) = 0$ implies the two distributions are equal
almost everywhere.  The divergence can be interpreted either in terms of codes
(see below), or in terms of statistical tests
\cite{Kullback-info-theory-and-stats}.  Roughly speaking, given $n$ samples
drawn from the distribution $\Dist$, the probability of our accepting the false
hypothesis that the distribution is $\AltDist$ can go down no faster than
$2^{-nD(\Dist\|\AltDist)}$.  The mutual information $I[X;Y]$ is the divergence
between the joint distribution $\Prob(X,Y)$, and the product of the marginal
distributions, $\Prob(X)\Prob(Y)$, and so measures the departure from
independence.

Some extra information-theoretic quantities make sense for time series and
stochastic processes.  Supposing we have a process $\bar{X} =$\\ $\ldots,
X_{-2}, X_{-1}, X_0, X_1, X_2, \ldots$, we can define its {\bold mutual
  information function} by analogy with the autocovariance function (see \S
\ref{sec:time-series-properties}),
\begin{eqnarray}
I_{\bar{X}}(s,t) & = & I[X_s;X_t]\\
I_{\bar{X}}(\tau) & = & I[X_t;X_{t+\tau}]
\end{eqnarray}
where the second form is valid only for strictly stationary processes.  The
mutual information function measures the degree to which different parts of
the series are dependent on each other.

The {\bold entropy rate} $h$ of a stochastic process
\begin{eqnarray}
h & \equiv & \lim_{L \rightarrow \infty}{H[X_0|X_{-L}^{-1}]}\\
& = & H[X_0|X_{-\infty}^{-1}] ~.
\end{eqnarray}
(The limit always exists for stationary processes.)  $h$ measures the process's
unpredictability, in the sense that it is the uncertainty which remains in the
next measurement even given complete knowledge of its past.  In nonlinear
dynamics, $h$ is called the {\bold Kolmogorov-Sinai (KS) entropy}.

For continuous variables, one can define the entropy via an integral,
\begin{eqnarray}
H[X] & \equiv & -\int{p(x) \log{p(x)} dx} ~,
\end{eqnarray}
with the subtlety that the continuous entropy not only can be negative, but
depends on the coordinate system used for $x$.  The relative entropy also has
the obvious definition,
\begin{eqnarray}
D(\Dist\|\AltDist) & \equiv & \int{p(x) \log{\frac{p(x)}{q(x)}} dx}
\end{eqnarray}
but is coordinate-independent and non-negative.  So, hence, is the mutual
information.

\paragraph{Optimal Coding}

One of the basic results of information theory concerns codes, or schemes for
representing random variables by bit strings.  That is, we want a scheme which
associates each value of a random variable $X$ with a bit string.  Clearly, if
we want to keep the average length of our code-words small, we should give
shorter codes to the more common values of $X$.  It turns out that the average
code-length is minimized if we use $-\log{\Prob(x)}$ bits to encode $x$, and it
is always possible to come within one bit of this.  Then, on average, we will
use $\Expec{-\log{\Prob(x)}} = H[X]$ bits.

This presumes we know the true probabilities.  If we think the true
distribution is $\AltDist$ when it is really $\Dist$, we will, on average, use
$\Expec{-\log{\AltDist(x)}} \geq H[X]$.  This quantity is called the {\bold
  cross-entropy} or {\bold inaccuracy}, and is equal to $H[X] +
D(\Dist\|\AltDist)$.  Thus, finding the correct probability distribution is
equivalent to minimizing the cross-entropy, or the relative entropy
\cite{Kulhavy-recursive}.

\paragraph{The Khinchin Axioms and R{\'e}nyi Information}

In 1953, A. I. Khinchin published a list of four reasonable-looking axioms for
a measure of the information $H[X]$ associated with a random variable $X$
\cite{Khinchin-info}.  He then proved that the Shannon information was the
unique functional satisfying the axioms, up to an over-all multiplicative
constant.  (The choice of this constant is equivalent to the choice of the base
for logarithms.) The axioms were as follows.
\begin{itemize}
\item The information is a functional of the probability distribution of $X$,
  and not on any of its other properties.  In particular, if $f$ is any
  invertible function, $H[X] = H[f(X)]$.
\item The information is maximal for the uniform distribution, where all events
  are equally probable.
\item The information is unchanged by enlarging the probability space with
  events of zero probability.
\item If the probability space is divided into two sub-spaces, so that $X$ is
  split into two variables $Y$ and $Z$, the total information is equal to the
  information content of the marginal distribution of one sub-space, plus the
  mean information of the conditional distribution of the other sub-space:
  $H[X] = H[Y] + \Expec{H(Z|Y)}$.
\end{itemize}
A similar axiomatic treatment can be given for the mutual information and the
relative entropy.

While the first three of Khinchin's axioms are all highly plausible, the fourth
is somewhat awkward.  It is intuitively more plausible to merely require that,
if $Y$ and $Z$ are independent, then $H[Y,Z] = H[Y] + H[Z]$.  If the fourth
axiom is weakened in this way, however, there is no longer only a single
functional satisfying the axioms.  Instead, any of the infinite family of
entropies introduced by R{\'e}nyi satisfies the axioms.  The {\bold R{\'e}nyi
  entropy} of order $\alpha$, with $\alpha$ any non-negative real number, is
\begin{eqnarray}
H_{\alpha}[X] & \equiv & \frac{1}{1 - \alpha}{\log{\sum_{i: p_i >
      0}{p_i^\alpha}}}
\end{eqnarray}
in the discrete case, and the corresponding integral in the continuous case.
The parameter $\alpha$ can be thought of as gauging how strongly the entropy is
biased towards low-probability events.  As $\alpha \rightarrow 0$,
low-probability events count more, until at $\alpha=0$, all possible events
receive equal weight.  (This is sometimes called the {\bold topological
  entropy}.)  As $\alpha\rightarrow\infty$, only the highest-probability event
contributes to the sum.  One can show that, as $\alpha\rightarrow 1$,
$H_{\alpha}[X] \rightarrow H[X]$, i.e., one recovers the ordinary Shannon
entropy in the limit.  There are entropy rates corresponding to all the
R{\'e}nyi entropies, defined just like the ordinary entropy rate.  For
dynamical systems, these are related to the fractal dimensions of the attractor
\cite{Ruelle-Lincei,Beck-Schlogl}.

The {\bold R{\'e}nyi divergences} bear the same relation to the R{\'e}nyi entropies as the Kullback-Leibler divergence does to the Shannon entropy.  The
defining formula is
\begin{eqnarray}
D_{\alpha}(\Dist||\AltDist) & \equiv & \frac{1}{\alpha-1}\log{\sum{q_i
      {\left(\frac{p_i}{q_i}\right)}^{\alpha}}}
\end{eqnarray}
and similarly for the continuous case.  Once again, $\lim_{\alpha\rightarrow
  1}{D_{\alpha}(\Dist||\AltDist)} = D(\Dist||\AltDist)$.  For all $\alpha > 0$,
$D_{\alpha}(\Dist||\AltDist) \geq 0$, and is equal to zero if and only if
$\Dist$ and $\AltDist$ are the same.  (If $\alpha = 0$, then a vanishing
R{\'e}nyi divergence only means that the supports of the two distributions are
the same.)  The R{\'e}nyi entropy $H_{\alpha}[X]$ is non-increasing as $\alpha$
grows, whereas the R{\'e}nyi divergence $D_{\alpha}(\Dist||\AltDist)$ is
non-decreasing.

\paragraph{Estimation of Information-Theoretic Quantities}

In applications, we will often want to estimate information-theoretic
quantities, such as the Shannon entropy or the mutual information, from
empirical or simulation data.  Restricting our attention, for the moment, to
the case of discrete-valued variables, the empirical distribution will
generally converge on the true distribution, and so the entropy (say) of the
empirical distribution (``sample entropy'') will also converge on the true
entropy.  However, it is not the case that the sample entropy is an {\em
  unbiased} estimate of the true entropy.  The Shannon (and R{\'e}nyi)
entropies are measures of variation, like the variance, and sampling tends to
reduce variation.  Just as the sample variance is a negatively biased estimate
of the true variance, sample entropy is a negatively-biased estimate of the
true entropy, and so sample mutual information is a positively-biased estimate
of true information.  Understanding and controlling the bias, as well as the
sampling fluctuations, can be very important.

Victor \cite{Victor-information-bias} has given an elegant method for
calculating the bias of the sample entropy; remarkably, the leading-order term
depends only on the alphabet size $k$ and the number of samples $N$, and is
$(k-1)/2N$.  Higher-order terms, however, depend on the true distribution.
Recently, Kraskov et al. \cite{Kraskov-et-al-estimating-MI} have published an
adaptive algorithm for estimating mutual information, which has very good
properties in terms of both bias and variance.  Finally, the estimation of
entropy {\em rates} is a somewhat tricky matter.  The best practices are to
either use an algorithm of the type given by
\cite{Kontoyiannis-et-al-estimating-entropy-rate}, or to fit a properly
dynamical model.  (For discrete data, variable-length Markov chains, discussed
in \S \ref{sec:VLMMs} above, generally work very well, and the entropy rate can
be calculated from them very simply.)  Another popular approach is to run one's
time series through a standard compression algorithm, such as \texttt{gzip},
dividing the size in bits of the output by the number of symbols in the input
\cite{Benedetto-et-al-language-trees}.  This is an absolutely horrible idea;
even under the circumstances under which it gives a consistent estimate of the
entropy rate, it converges much more slowly, and runs more slowly, than
employing either of the two techniques just mentioned
\cite{Khmelev-Teahan-on-Benedetto-et-al,Goodman-on-Benedetto-et-al}.\footnote{For
  a pedagogical discussion, with examples, of how compression algorithms may be
  misused, see \url{http://bactra.org/notebooks/cep-gzip.html}.}

\subsection{Applications of Information Theory}

Beyond its original home in communications engineering, information theory has
found a multitude of applications in statistics
\cite{Kullback-info-theory-and-stats,Kulhavy-recursive} and learning theory
\cite{Jordan-learning-in-graphical-models,MacKay-learning-algorithms}.
Scientifically, it is very natural to consider some biological systems as
communications channels, and so analyze their information content; this has
been particularly successful for biopolymer sequences
\cite{Bio-sequence-analysis} and especially for neural systems, where the
analysis of neural codes depends vitally on information theory
\cite{Spikes-book,Abbott-Sejnowski-neural-codes}.  However, there is nothing
prohibiting the application of information theory to systems which are not
designed to function as communications devices; the concepts involved require
only well-defined probability distributions.  For instance, in nonlinear
dynamics \cite{Billingsley-ergodic-theory-and-info,Katok-Hasselblatt}
information-theoretic notions are very important in characterizing different
kinds of dynamical system (see also \S \ref{sec:symbolic-dyn}).  Even more
closely tied to complex systems science is the literature on ``physics and
information'' or ``physics and computation'', which investigates the
relationships between the mechanical principles of physics and information
theory, e.g., Landauer's principle, that erasing (but not storing) a bit of
information at temperature $T$ produces $k_B T \ln{2}$ joules of heat, where
$k_B$ is Boltzmann's constant.

\section{Complexity Measures}
\label{sec:complexity}

We have already given some thought to complexity, both in our initial rough
definition of ``complex system'' and in our consideration of machine learning
and Occam's Razor.  In the latter, we saw that the relevant sense of
``complexity'' has to do with families of models: a model class is complex if
it requires large amounts of data to reliably find the best model in the class.
On the other hand, we initially said that a complex system is one with many
highly variable, strongly interdependent parts.  Here, we will consider various
proposals for putting some mathematical spine into that notion of a system's
complexity, as well as the relationship to the notion of complexity of
learning.

Most measures of complexity for systems formalize the intuition that something
is complex if it is difficult to describe adequately.  The first mathematical
theory based on this idea was proposed by Kolmogorov; while it is {\em not}
good for analyzing empirical complex systems, it was very important
historically, and makes a good point of entry into the field.

\subsection{Algorithmic Complexity}

Consider a collection of measured data-values, stored in digitized form on a
computer.  We would like to say that they are complex if they are hard to
describe, and measure their complexity by the difficulty of describing them.
The central idea of Kolmogorov complexity (proposed independently by Solomonoff
\cite{Solomonoff} and Chaitin) is that one can describe the data set by writing
a program which will reproduce the data.  The difficulty of description is then
measured by the length of the program.  Anyone with much experience of other
people's code will appreciate that it is always possible to write a longer,
slower program to do a given job, so what we are really interested in is the
shortest program that can exactly replicate the data.

To introduce some symbols, let $x$ be the data, and $|x|$ their size in bits.
The Kolmogorov or algorithmic complexity of $x$, $K(x)$, is the length of the
shortest program that will output $x$ and then stop\footnote{The issue of what
  language to write the program in is secondary; writing a program to convert
  from one language to another just adds on a constant to the length of the
  over-all program, and we will shortly see why additive constants are not
  important here.}.  Clearly, there is always some program which will output
$x$ and then stop, for instance, ``\texttt{print($x$); end}''.  Thus $K(x) \leq
|x| + c$, where $c$ is the length of the print and end instructions.  This is
what one might call a literal description of the data.  If one cannot do better
than this --- if $K(x) \approx |x|$ --- then $x$ is highly complex.  Some data,
however, is highly compressible.  For instance, if $x$ consists of the second
quadrillion digits of $\pi$, a very short program suffices to generate
it\footnote{Very short programs can calculate $\pi$ to arbitrary accuracy, and
  the length of these programs does not grow as the number of digits calculated
  does.  So one could run one of these programs until it had produced the first
  two quadrillion digits, and then erase the first half of the output, and
  stop.}.

As you may already suspect, the number of simple data sets is quite limited.
Suppose we have a data set of size $n$ bits, and we want to compress it by $k$
bits, i.e., find a program for it which is $n-k$ bits long.  There are at most
$2^{n-k}$ programs of that length, so of all the $2^n$ data sets of size $n$,
the fraction which can be compressed by $k$ bits is at most $2^{-k}$.  The
precise degree of compression does not matter --- when we look at large data
sets, almost all of them are highly complex.  If we pick a large data set {\em
  at random}, then the odds are very good that it will be complex.  We can
state this more exactly if we think about our data as consisting of the first
$n$ measurements from some sequence, and let $n$ grow.  That is, $x = x_1^n$,
and we are interested in the asymptotic behavior of $K(x_1^n)$.  If the
measurements $x_i$ are independent and identically distributed (IID), then
$K(x_1^n)/|x| \rightarrow 1$ almost surely; IID sequences are {\bold
  incompressible}.  If $x$ is a realization of a stationary (but not
necessarily IID) random process $\bar{X}$, then
\cite{Li-and-Vitanyi-1993,Badii-Politi}
\begin{eqnarray}
\lim_{n\rightarrow\infty}{\Expec{\frac{K(X_1^n)}{n}}} & = & h(\bar{X}) ~,
\end{eqnarray}
the entropy rate (\S \ref{sec:info-theory}) of $\bar{X}$.  Thus, random data
has high complexity, and the complexity of a random process grows at a rate
which just measures its unpredictability.

This observation goes the other way: complex data looks random.  The heuristic
idea is that if there were any regularities in the data, we could use them to
shave at least a little bit off the length of the minimal program.  What one
can show formally is that incompressible sequences have {\em all} the
properties of IID sequences --- they obey the law of large numbers and the
central limit theorem, pass all statistical tests for randomness, etc.  In
fact, this possibility, of defining ``random'' as ``incompressible'', is what
originally motivated Kolmogorov's work \cite[chapter 3]{Salmon-1984}.

Kolmogorov complexity is thus a very important notion for the foundations of
probability theory, and it has extensive applications in theoretical computer
science \cite{Li-and-Vitanyi-1993} and even some aspects of statistical physics
\cite{Zurek-demon-of-choice}.  Unfortunately, it is quite useless as a measure
of the complexity of natural systems.  This is for two reasons.  First, as we
have just seen, it is maximized by {\em independent} random variables; we want
{\em strong dependence}.  Second, and perhaps more fundamental, it is simply
not possible to calculate Kolmogorov complexity.  For deep reasons related to
G{\"o}del's Theorem, there cannot be any procedure for calculating $K(x)$, nor
are there any accurate approximation procedures \cite{Li-and-Vitanyi-1993}.

Many scientists are strangely in denial about the Kolmogorov complexity, in
that they think they can calculate it.  Apparently unaware of the mathematical
results, but aware of the relationship between Kolmogorov complexity and data
compression, they reason that file compression utilities should provide an
estimate of the algorithmic information content.  Thus one finds many papers
which might be titled ``\texttt{gzip} as a measure of
complexity''\footnote{\cite{Benedetto-et-al-language-trees} is perhaps the most
  notorious; see \cite{Khmelev-Teahan-on-Benedetto-et-al} and especially
  \cite{Goodman-on-Benedetto-et-al} for critiques.}, and the practice is even
recommended by some otherwise-reliable sources (e.g.,
\cite{Sprott-on-time-series}).  However, this is simply a confused idea, with
absolutely nothing to be said in its defense.

\subsection{Refinements of Algorithmic Complexity}

We saw just now that algorithmic information is really a measure of randomness,
and that it is maximized by collections of independent random variables.  Since
complex systems have many strongly dependent variables, it follows that the
Kolmogorov notion is not the one we really want to measure.  It has long been
recognized that we really want something which is small both for systems which
are strongly ordered (i.e., have only a small range of allowable behavior) and
for those which are strongly disordered (i.e., have independent parts).  Many
ways of modifying the algorithmic information to achieve this have been
proposed; two of them are especially noteworthy.

\subsubsection{Logical Depth}
Bennett \cite{Bennett-dissipation,Bennett-1986,Bennett-how-and-why} proposed
the notion of the {\bold logical depth} of data as a measure of its complexity.
Roughly speaking, the logical depth $L(x)$ of $x$ is the number of
computational steps the minimal program for $x$ must execute.  For
incompressible data, the minimal program is \texttt{print($x$)}, so $L(x)
\approx |x|$.  For periodic data, the minimal program cycles over printing out
one period over and over, so $L(x) \approx |x|$ again.  For some compressible
data, however, the minimal program must do non-trivial computations, which are
time-consuming.  Thus, to produce the second quadrillion digits of $\pi$, the
minimal program is one which {\em calculates} the digits, and this takes
considerably more time than reading them out of a list.  Thus, $\pi$ is deep,
while random or periodic data are shallow.

While logical depth is a clever and appealing idea, it suffers from a number of
drawbacks.  First, real data are not, so far as we know, actually produced by
running their minimal programs\footnote{It is certainly legitimate to regard
  any dynamical process as also a computational process,
  \cite{Churchland-Sejnowski,Giunti-computation-dynamics-cognition,%
    Margolus-crystalline,CMPPSS}, so one could argue that the data {\em is}
  produced by some kind of program.  But even so, this computational process
  generally does not resemble that of the minimal, Kolmogorov program at all.},
and the run-time of that program has no known {\em physical} significance, and
that's not for lack of attempts to find one \cite{Lloyd-Pagels}.  Second, and
perhaps more decisively, there is still no procedure for finding the minimal
program.

\subsubsection{Algorithmic Statistics}

Perhaps the most important modification of the Kolmogorov complexity is that
proposed by G{\'a}cs, Tromp and Vitanyi \cite{Gacs-Tromp-Vitanyi}, under the
label of ``algorithmic statistics''.  Observe that, when speaking of the
minimal program for $x$, I said nothing about the inputs to the program; these
are to be built in to the code.  It is this which accounts for the length of
the programs needed to generate random sequences: almost all of the length of
\texttt{print($x$)} comes from $x$, not $\texttt{print()}$.  This suggests
splitting the minimal program into two components, a ``model'' part, the
program properly speaking, and an ``data'' part, the inputs to the program.  We
want to put all the regularities in $x$ into the model, and all the arbitrary,
noisy parts of $x$ into the inputs.  Just as in probability theory a
``statistic'' is a function of the data which summarizes the information they
convey, G{\'a}cs {\em et al.} regard the model part of the program as an {\bold
  algorithmic statistic}, summarizing its regularities.  To avoid the trivial
regularity of $\texttt{print()}$ when possible, they define a notion of a
{\bold sufficient} algorithmic statistic, based on the idea that $x$ should be
in some sense a typical output of the model (see their paper for details).
They then define the complexity of $x$, or, as they prefer to call it, the
{\bold sophistication}, as the length of the shortest sufficient algorithmic
statistic.

Like logical depth, sophistication is supposed to discount the purely random
part of algorithmic complexity.  Unlike logical depth, it stays within the
confines of description in doing so; programs, here, are just a particular,
mathematically tractable, kind of description.  Unfortunately, the
sophistication is still uncomputable, so there is no real way of applying
algorithmic statistics.

\subsection{Statistical Measures of Complexity}
\label{sec:stat-compl}

The basic problem with algorithmic complexity and its extensions is that they
are all about finding the shortest way of exactly describing a single
configuration.  Even if we could compute these measures, we might suspect, on
the basis of our discussion of over-fitting in \S \ref{sec:data-mining} above,
that this is not what we want.  Many of the details of any particular set of
data are just noise, and will not generalize to other data sets obtained from
the same system.  If we want to characterize the complexity of the system, it
is precisely the generalizations that we want, and not the noisy particulars.
Looking at the sophistication, we saw the idea of picking out, from the overall
description, the part which describes the regularities of the data.  This idea
becomes useful and operational when we abandon the goal of {\em exact}
description, and allow ourselves to recognize that the world is full of noise,
which is easy to describe statistically; we want a statistical, and not an
algorithmic, measure of complexity.

I will begin with what is undoubtedly the most widely-used statistical measure
of complexity, Rissanen's {\bold stochastic complexity}, which can also be
considered a method of model selection.  Then I will look at three attempts to
isolate the complexity of the system as such, by considering how much
information would be required to predict its behavior, {\em if} we had an
optimal statistical model of the system.

\subsubsection{Stochastic Complexity and the Minimum Description Length}
\label{sec:mdl}

Suppose we have a statistical model with some parameter $\theta$, and we
observe the data $x$.  The model assigns a certain likelihood to the data,
$\Prob_{\theta}(X=x)$.  One can make this into a loss function by taking its
negative logarithm: $L(\theta,x) = -\log{\Prob_{\theta}(X=x)}$.  Maximum
likelihood estimation minimizes this loss function.  We also learned, in \S
\ref{sec:info-theory}, that if $\Prob_{\theta}$ is the correct probability
distribution, the optimal coding scheme will use $-\log{\Prob_{\theta}(X=x)}$
bits to encode $x$.  Thus, maximizing the likelihood can also be thought of as
minimizing the encoded length of the data.

However, we do not yet have a complete description: we have an encoded version
of the data, but we have not said what the encoding scheme, i.e., the model,
is.  Thus, the total description length has two parts:
\begin{eqnarray}
C(x, \theta, \Theta) & = & L(x,\theta) + D(\theta, \Theta)
\end{eqnarray}
where $D(\theta, \Theta)$ is the number of bits we need to specify $\theta$
from among the set of all our models $\Theta$.  $L(x,\theta)$ represents the
``noisy'' or arbitrary part of the description, the one which will not
generalize; the model represents the part which does generalize.  If $D(\theta,
\Theta)$ gives short codes to simple models, we have the desired kind of
trade-off, where we can reduce the part of the data which looks like noise only
by using a more elaborate model.  The {\bold minimum description length
  principle} \cite{Rissanen-1978,Rissanen-SCiSI} enjoins us to pick the model
which minimizes the description length, and the {\bold stochastic complexity}
of the data is that minimized description-length:
\begin{eqnarray}
{\theta}_{\mathrm{MDL}} & = & \argmin{\theta}{C(x, \theta, \Theta)}\\
C_{\mathrm{SC}}(x, \Theta) & = & \min_{\theta}{C(x, \theta, \Theta)}
\end{eqnarray}
Under not-too-onerous conditions on the underlying data-generating process and
the model class $\Theta$ \cite[chapter 3]{Rissanen-SCiSI}, as we provide more
data $\theta_{\mathrm{MDL}}$ will converge on the model in $\Theta$ which
minimizes the generalization error, which here is just the same as minimizing
the Kullback-Leibler divergence from the true distribution\footnote{It is
  important to note \cite[chapter 3]{Rissanen-SCiSI} that if we allowed any
  possible model in $\Theta$, find the minimum would, once again, be
  incomputable.  This restriction to a definite, perhaps hierarchically
  organized, class of models is vitally important.}.

Regarded as a principle of model selection, MDL has proved very successful in
many applications, even when dealing with quite intricate,
hierarchically-layered model classes.  (\cite{Hraber-et-al-MDL-and-HIV} is a
nice recent application to a biomedical complex system; see \S
\ref{sec:nld-approach} for applications to state-space reconstruction.)  It is
important to recognize, however, that most of this success comes from carefully
tuning the model-coding term $D(\theta,\Theta)$ so that models which do not
generalize well turn out to have long encodings.  This is perfectly legitimate,
but it relies on the tact and judgment of the scientist, and often, in dealing
with a complex system, we have no idea, or at least no {\em good} idea, what
generalizes and what does not.  If we were malicious, or short-sighted, we can
always insure that the particular data we got have a stochastic complexity of
just one bit\footnote{Take our favorite class of models, and add on
  deterministic models which produce particular fixed blocks of data with
  probability 1.  For any of these models $\theta$, $L(x,\theta)$ is either 0
  (if $x$ is what that model happens to generate) or $\infty$.  Then, once we
  have our data, and find a $\theta$ which generates that and nothing but that,
  re-arrange the coding scheme so that $D(\theta,\Theta) = 1$; this is always
  possible.  Thus, $C_{\mathrm{SC}}(x,\Theta) = 1$ bit.}.  The model which
gives us this complexity will then have absolutely horrible generalization
properties\footnote{This does not contradict the convergence result of the last
  paragraph; one of the not-too-onerous conditions mentioned in the previous
  paragraph is that the coding scheme remain fixed, and we're violating that.}.

Whatever its merits as a model selection method, stochastic complexity does not
make a good measure of the complexity of natural systems.  There are at least
three reasons for this.
\begin{enumerate}
\item The dependence on the model-encoding scheme, already discussed.
\item The log-likelihood term, $L(x,\theta)$ in $C_{\mathrm{SC}}$ can be
  decomposed into two parts, one of which is related to the entropy rate of the
  data-generating process, and so reflects its intrinsic unpredictability.  The
  other, however, indicates the degree to which even the most accurate model in
  $\Theta$ is misspecified.  Thus it reflects our ineptness as modelers, rather
  than any characteristic of the process.
\item Finally, the stochastic complexity reflects the need to specify some
  particular model, and to represent this specification.  While this is
  necessarily a part of the modeling process for us, it seems to have no {\em
    physical} significance; the system does not need to {\em represent} its
  organization, it just {\em has} it.
\end{enumerate}

\subsubsection{Complexity via Prediction}
\label{sec:comp-mech}

\paragraph{Forecast Complexity and Predictive Information}

Motivated in part by concerns such as these, Grassberger
\cite{Grassberger-1986} suggested a new and highly satisfactory approach to
system complexity: complexity is the amount of information required for optimal
prediction.  Let us first see why this idea is plausible, and then see how it
can be implemented in practice.  (My argument does not follow that of
Grassberger particularly closely.  Also, while I confine myself to time series,
for clarity, the argument generalizes to any kind of prediction
\cite{CRS-thesis}.)

We have seen that there is a limit on the accuracy of any prediction of a given
system, set by the characteristics of the system itself (limited precision of
measurement, sensitive dependence on initial conditions, etc.).  Suppose we had
a model which was maximally predictive, i.e., its predictions were at this
limit of accuracy.  Prediction, as I said, is always a matter of mapping inputs
to outputs; here the inputs are the previous values of the time series.
However, not all aspects of the entire past are relevant.  In the extreme case
of independent, identically-distributed values, {\em no} aspects of the past
are relevant.  In the case of periodic sequences with period $p$, one only
needs to know which of the $p$ phases the sequence is in.  If we ask how {\em
  much} information about the past is relevant in these two cases, the answers
are clearly 0 and $\log{p}$, respectively.  If one is dealing with a Markov
chain, only the present state is relevant, so the amount of information needed
for optimal prediction is just equal to the amount of information needed to
specify the current state.  One thus has the nice feeling that both highly
random (IID) and highly ordered (low-period deterministic) sequences are of low
complexity, and more interesting cases can get high scores.

More formally, any predictor $f$ will translate the past of the sequence $x^-$
into an effective state, $s = f(x^-)$, and then make its prediction on the
basis of $s$.  (This is true whether $f$ is formally a state-space model or
not.)  The amount of information required to specify the state is $H[S]$.  We
can take this to be the complexity of $f$.  Now, if we confine our attention to
the set $\mathcal{M}$ of maximally predictive models, we can define what
Grassberger called the ``true measure complexity'' or ``forecast complexity''
of the process as the minimal amount of information needed for optimal
prediction:
\begin{eqnarray}
C & = & \min_{f\in\mathcal{M}}{H[f({X}^{-})]}
\end{eqnarray}

Grassberger did not provide a procedure for finding the maximally predictive
models, nor for minimizing the information required among them.  He did,
however, make the following observation.  A basic result of information theory,
called the {\bold data-processing inequality}, says that $I[A;B] \geq
I[f(A);B]$, for any variables $A$ and $B$ --- we cannot get more information
out of data by processing it than was in there to begin with.  Since the state
of the predictor is a function of the past, it follows that $I[X^-;X^+] \geq
I[f(X^-);X^+]$.  Presumably, for optimal predictors, the two informations are
equal --- the predictor's state is just as informative as the original data.
(Otherwise, the model would be missing some potential predictive power.)  But
another basic inequality is that $H[A] \geq I[A;B]$ --- no variable contains
more information about another than it does about itself.  So, for optimal
models, $H[f(X^-)] \geq I[X^-;X^+]$.  The latter quantity, which Grassberger
called the {\bold effective measure complexity}, can be estimated purely from
data, without intervening models.  This quantity --- the mutual information
between the past and the future --- has been rediscovered many times, in many
contexts, and called {\bold excess entropy} (in statistical mechanics), {\bold
  stored information} \cite{Shaw-dripping}, {\bold complexity}
\cite{Lindgren-Nordahl-1988,Li-complexity-vs-entropy,%
  Arnold-info-theory-phase-trans} or {\bold predictive information}
\cite{Bialek-Nemenman-Tishby}; the last name is perhaps the clearest.  Since it
quantifies the degree of statistical dependence between the past and the
future, it is clearly appealing as a measure of complexity.

\paragraph{The Grassberger-Crutchfield-Young Statistical Complexity}

The forecasting complexity notion was made fully operational by Crutchfield and
Young \cite{Inferring-stat-compl,Computation-at-the-onset}, who provided an
effective procedure for finding the minimal maximally predictive model and its
states.  They began by defining the {\bold causal states} of a process, as
follows.  For each history $x^-$, there is some conditional distribution of
future observations, $\Prob(X^+|x^-)$.  Two histories $x_1^-$ and $x_2^-$ are
equivalent if $\Prob(X^+|x_1^-) = \Prob(X^+|x_2^-)$.  Write the set of all
histories equivalent to $x^-$ as $[x^-]$.  Now we have a function $\epsilon$
which maps each history into its equivalence class: $\epsilon(x^-) = [x^-]$.
Clearly, $\Prob(X^+|\epsilon(x^-)) = \Prob(X^+|x^-)$.  Crutchfield and Young
accordingly proposed to forget the particular history and retain only its
equivalence class, which they claimed would involve no loss of predictive
power; this was later proved to be correct \cite[theorem 1]{CMPPSS}.  They
called the equivalence classes the ``causal states'' of the process, and
claimed that these were the simplest states with maximal predictive power; this
is also was right \cite[theorem 2]{CMPPSS}.  Finally, one can show that the
causal states are the {\em unique} optimal states \cite[theorem 3]{CMPPSS}; any
other optimal predictor is really a disguised version of the causal states.
Accordingly, they defined the {\bold statistical complexity} of a process $C$
as the information content of its causal states.

Because the causal states are purely an objective property of the process being
considered, $C$ is too; it does not depend at all on our modeling or means of
description.  It is equal to the length of the shortest description of the past
which is {\em relevant} to the actual dynamics of the system.  As we argued
should be the case above, for IID sequences it is exactly 0, and for periodic
sequences it is $\log{p}$.  One can show \cite[theorems 5 and 6]{CMPPSS} that
the statistical complexity is always at least as large as the predictive
information, and generally that it measures how far the system departs from
statistical independence.

The causal states have, from a statistical point of view, quite a number of
desirable properties.  The maximal prediction property corresponds exactly to
that of being a sufficient statistic \cite{Kullback-info-theory-and-stats}; in
fact they are minimal sufficient statistics
\cite{Kullback-info-theory-and-stats,CMPPSS}.  The sequence of states of the
process form a Markov chain.  Referring back to our discussion of filtering and
state estimation (\S \ref{sec:filtering}), one can design a recursive filter
which will eventually estimate the causal state without any error at all;
moreover, it is always clear whether the filter has ``locked on'' to the
correct state or not.

All of these properties of the causal states and the statistical complexity
extend naturally to spatially-extended systems, including, but not limited to,
cellular automata \cite{CRS-prediction-on-networks,QSO-in-PRL}.  Each point in
space then has its own set of causal states, which form not a Markov chain but
a Markov field, and the local causal state is the minimal sufficient statistic
for predicting the future of that point.  The recursion properties carry over,
not just temporally but spatially: the state at one point, at one time, helps
determine not only the state at that same point at later times, but also the
state at neighboring points at the same time.  The statistical complexity, in
these spatial systems, becomes the amount of information needed about the past
of a given point in order to optimally predict its future.  Systems with a high
degree of local statistical complexity are ones with intricate spatio-temporal
organization, and, experimentally, increasing statistical complexity gives a
precise formalization of intuitive notions of self-organization
\cite{QSO-in-PRL}.

Crutchfield and Young were inspired by analogies to the theory of abstract
automata, which led them to call their theory, somewhat confusingly, {\bold
  computational mechanics}.  Their specific initial claims for the causal
states were based on a procedure for deriving the minimal automaton capable of
producing a given family of sequences\footnote{Technically, a given regular
  language (\S \ref{sec:symbolic-dyn}).} known as Nerode equivalence classing
\cite{Lewis-Papadimitriou-computation}.  In addition to the theoretical
development, the analogy to Nerode equivalence-classing led them to describe a
procedure \cite{Inferring-stat-compl} for estimating the causal states and the
$\epsilon$-machine from empirical data, at least in the case of discrete
sequences.  This Crutchfield-Young algorithm has actually been successfully
used to analyze empirical data, for instance, geomagnetic fluctuations
\cite{Watkins-et-al-comp-mech-of-geomag}.  The algorithm has, however, been
superseded by a newer algorithm which uses the known properties of the causal
states to guide the model discovery process \cite{CSSR-for-UAI} (see \S
\ref{sec:causal-state-models} above).

Let me sum up.  The Grassberger-Crutchfield-Young statistical complexity is an
objective property of the system being studied.  It reflects the {\em
  intrinsic} difficulty of predicting it, namely the amount of information
which is actually relevant to the system's dynamics.  It is low both for highly
disordered and trivially-ordered systems.  Above all, it is calculable, and has
actually been calculated for a range of natural and mathematical systems.
While the initial formulation was entirely in terms of discrete time series,
the theory can be extended straightforwardly to spatially-extended dynamical
systems \cite{CRS-prediction-on-networks}, where it quantifies
self-organization \cite{QSO-in-PRL}, to controlled dynamical systems and
transducers, and to prediction problems generally \cite{CRS-thesis}.

\subsection{Power Law Distributions}

Over the last decade or so, it has become reasonably common to see people
(especially physicists) claiming that some system or other is complex, because
it exhibits a power law distribution of event sizes.  Despite its popularity,
this is simply a fallacy.  No one has demonstrated any relation between power
laws and any kind of formal complexity measure.  Nor is there any link tying
power laws to our intuitive idea of complex systems as ones with strongly
interdependent parts.

It is true that, {\em in equilibrium statistical mechanics}, one does not find
power laws {\em except} near phase transitions
\cite{Chandler-modern-stat-mech}, when the system {\em is} complex by our
standard.  This has encouraged physicists to equate power laws as such with
complexity.  Despite this, it has been known for half a century
\cite{Simon-skew} that there are many, many ways of generating power laws, just
as there are many mechanisms which can produce Poisson distributions, or
Gaussians.  Perhaps the simplest one is that recently demonstrated by Reed and
Hughes \cite{Reed-and-Hughes-on-power-laws}, namely exponential growth coupled
with random observation times.  The observation of power laws alone thus says
nothing about complexity (except in thermodynamic equilibrium!), and certainly
is not a reliable sign of some specific favored mechanism, such as
self-organized criticality \cite{Bak-Tang-and-Wiesenfeld,Jensen-on-SOC} or
highly-optimized tolerance
\cite{Carlson-Doyle-HOT-1,Carlson-Doyle-HOT-2,MEJN-Girvan-COLD}.

\subsubsection{Statistical Issues Relating to Power Laws}

The statistics of power laws are poorly understood within the field of complex
systems, to a degree which is quite surprising considering how much attention
has been paid to them. To be quite honest, there is little reason to think that
many of the things claimed to be power laws actually {\em are} such, as opposed
to some other kind of heavy-tailed distribution.  This brief section will
attempt to inoculate the reader against some common mistakes, most of which
are related to the fact that a power law makes a straight line on a log-log
plot.  Since it would be impractical to cite all papers which commit these
mistakes, and unfair to cite only some of them I will omit references here;
interested readers will be able to assemble collections of their own very
rapidly.

\paragraph{Parameter Estimation}
Presuming that something is a power law, a natural way of estimating its
exponent is to use linear regression to find the line of best fit to the points
on the log-log plot.  This is actually a consistent estimator, if the data
really do come from a power law.  However, the loss function used in linear
regression is the sum of the squared distances between the line and the points
(``least squares'').  In general, the line minimizing the sum of squared errors
is {\em not} a valid probability distribution, and so this is simply not a
reliable way to estimate the {\em distribution}.

One is much better off using maximum likelihood to estimate the parameter.
With a discrete variable, the probability function is $\Prob(X=x) =
x^{-\alpha}/\zeta(\alpha)$, where $\zeta(\alpha) =
\sum_{k=1}^{\infty}{k^{-\alpha}}$ is the Riemann zeta function, which ensures
that the probability is normalized.  Thus the maximum likelihood estimate of
the exponent is obtained by minimizing the negative log-likelihood, $L(\alpha)
= \sum_{i}{\alpha \log{x_i}} +\log{\zeta(\alpha)}$, i.e., $L(\alpha)$ is our
loss function.  In the continuous case, the probability density is $(\alpha-1)
c^{\alpha-1}/x^{\alpha}$, with $x \geq c > 0$.

\paragraph{Error Estimation}
Most programs to do linear regression also provide an estimate of the standard
error in the estimated slope, and one sometimes sees this reported as the
uncertainty in the power law.  This is an entirely unacceptable procedure.
Those calculations of the standard error assume that measured values having
Gaussian fluctuations around their true means.  Here that would mean that the
log of the empirical relative frequency is equal to the log of the probability
plus Gaussian noise.  However, by the central limit theorem, one knows that the
relative frequency is equal to the probability plus Gaussian noise, so the
former condition does not hold.  Notice that one can obtain asymptotically
reliable standard errors from maximum likelihood estimation.

\paragraph{Validation; $R^2$}
Perhaps the most pernicious error is that of trying to validate the assumption
of a power law distribution by either eye-balling the fit to a straight line,
or evaluating it using the $R^2$ statistic, i.e., the fraction of the variance
accounted for by the least-squares regression line.  Unfortunately, while these
procedures are good at confirming that something is a power law, if it really
is (low Type I error, or high statistical significance), they are very bad at
alerting you to things that are {\em not} power laws (they have a very high
rate of Type II error, or low statistical power).  The basic problem here is
that {\em any} smooth curve looks like a straight line, if you confine your
attention to a sufficiently small region --- and for some non-power-law
distributions, such ``sufficiently small'' regions can extend over multiple
orders of magnitude.

\begin{figure}[t]
\begin{center}
\resizebox{3in}{!}{\includegraphics{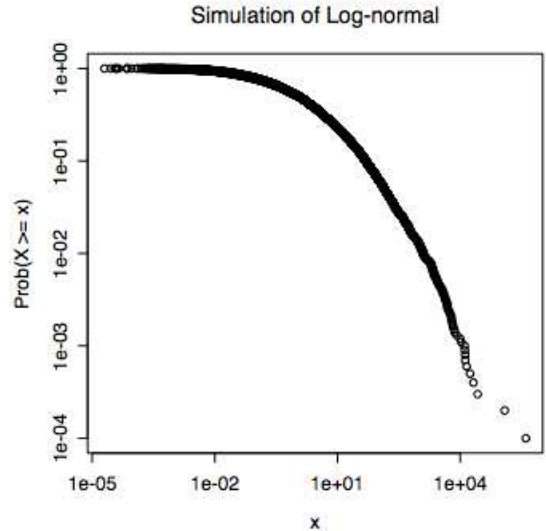}}
\end{center}
\caption{\label{fig:simulation} Distribution of 10,000 random numbers,
  generated according to a log-normal distribution with $\Expec{\log{X}} = 0$
  and $\sigma(\log{X}) = 3$.}
\end{figure}

\begin{figure}[t]
\begin{center}
\resizebox{3in}{!}{\includegraphics{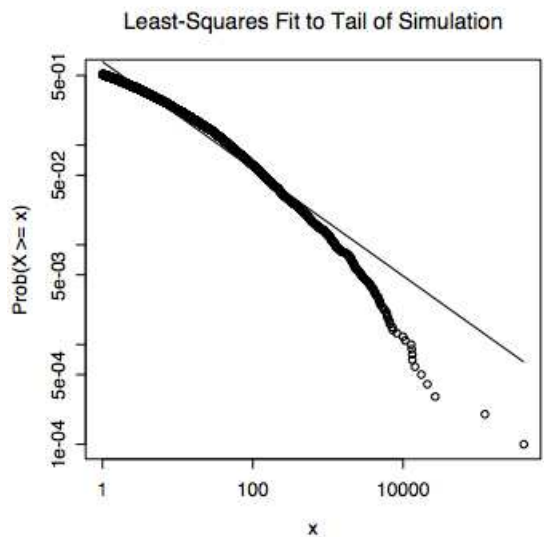}}
\end{center}
\caption{\label{fig:simulation-tail} Inability of linear regression on log-log
  plots to correctly identify power law distributions.  Simulation data
  (circles) and resulting least-squares fit (line) for the 5,112 points in
  Figure \ref{fig:simulation} for which $x \geq 1$.  The $R^2$ of the
  regression line is 0.962.}
\end{figure}

To illustrate this last point, consider Figure \ref{fig:simulation}, made by
generating 10,000 random numbers according to a log-normal distribution, with a
mean log of 0 and a standard deviation in the log of 3.
Restricting attention to the ``tail'' of random numbers $\geq 1$, and
doing a usual least-squares fit, gives the line shown in Figure \ref{fig:simulation-tail}.
One might hope that it would be easy to tell that this data does not come from
a power law, since there are a rather large number of observations (5,112),
extending over a wide domain (more than four orders of magnitude).
Nonetheless, $R^2$ is 0.962.  This, in and of itself, constitutes a
demonstration that getting a high $R^2$ is not a reliable indicator that one's
data was generated by a power law.\footnote{If I replace the random data by the
  {\em exact} log-normal probability distribution over the same range, and do a
  least-squares fit to that, the $R^2$ actually increases, to 0.994.}

\begin{figure}[t]
\begin{center}
\resizebox{3in}{!}{\includegraphics{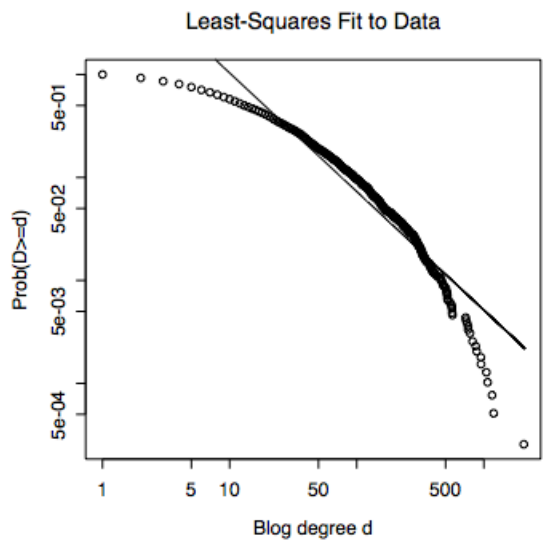}}
\end{center}
\caption{\label{fig:least-squares-fit} Empirical distribution of the in-degrees
  of political weblogs (``blogs'').  Horizontal axis: number of incoming links
  $d$; vertical axis: fraction of all blogs with at least that many links,
  $\Prob(D \geq d)$; both axes are on a log-log scale.  Circles show the actual
  distribution; the straight line is a least-squares fit to these values.  This
  does not produce a properly normalized probability distribution but it does
  have an $R^2$ of 0.898, despite the clear concavity of the curve.}
\end{figure}

\paragraph{An Illustration: Blogging}
An amusing empirical illustration of the difficulty of distinguishing between
power laws and other heavy-tailed distributions is provided by political
weblogs, or ``blogs'' --- websites run by individuals or small groups providing
links and commentary on news, political events, and the writings of other
blogs.  A rough indication of the prominence of a blog is provided by the
number of other blogs linking to it --- its {\bold in-degree}.  (For more on
network terminology, see Wuchty, Ravasz and Barab{\'a}si, this volume.)  A
widely-read essay by Shirky claimed that the distribution of in-degree follows
a power law, and used that fact, and the literature on the growth of scale-free
networks, to draw a number of conclusions about the social organization of the
blogging community \cite{Shirky-power-law}.  A more recent paper by Drenzer and
Farrell \cite{Drenzer-Farrell}, in the course of studying the role played by
blogs in general political debate, re-examined the supposed power-law
distribution.\footnote{Profs.\ Drenzer and Farrell kindly shared their data
  with me, but the figures and analysis that follow are my own.}  Using a large
population of inter-connected blogs, they found a definitely heavy-tailed
distribution which, on a log-log plot, was quite noticeably concave
(\ref{fig:least-squares-fit}); nonetheless, $R^2$ for the conventional
regression line was 0.898.

\begin{figure}[t]
\begin{center}
\resizebox{3in}{!}{\includegraphics{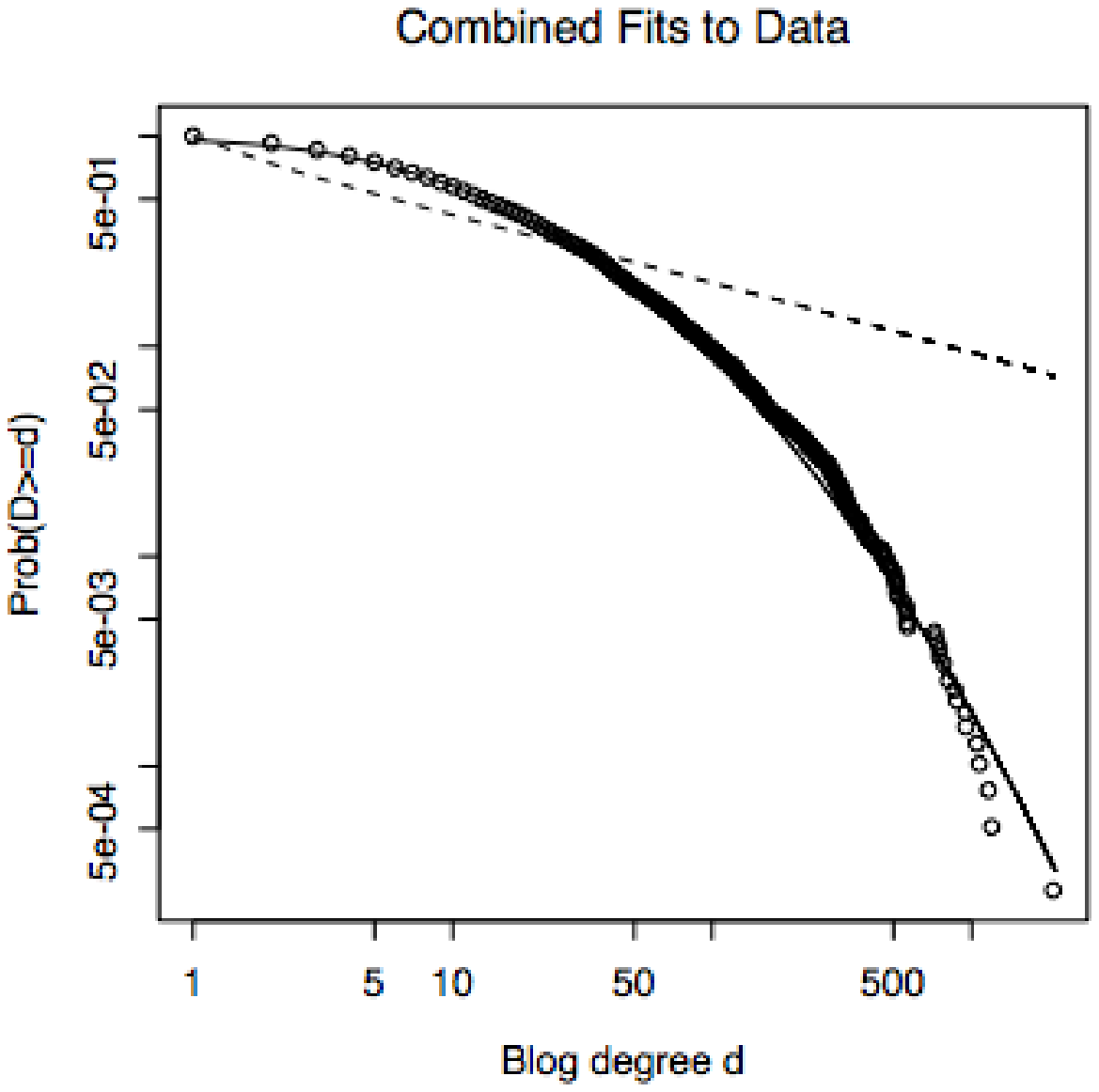}}
\end{center}
\caption{\label{fig:combined-fit} Maximum likelihood fits of log-normal (solid
  line) and power law (dashed line) distributions to the data from Figure
  \ref{fig:least-squares-fit} (circles); axes as in that figure.  Note the
  extremely tight fit of the log-normal over the whole range of the curve, and
  the general failure of the power-law distribution.}
\end{figure}

Maximum likelihood fitting of a power law distribution gave $\alpha = -1.30 \pm
0.006$, with a negative log-likelihood of $18481.51$.  Similarly fitting a
log-normal distribution gave $\Expec{\log{X}} =2.60 \pm 0.02$ and
$\sigma(\log{X}) = 1.48 \pm 0.02$, with a negative log-likelihood of 17,218.22.
As one can see from Figure \ref{fig:combined-fit}, the log-normal provides a
very good fit to almost all of the data, whereas even the best fitting
power-law distribution is not very good at all.\footnote{Note that the
  log-normal curve fitted to the {\em whole} data continues to match the data
  well even in the tail.  For further discussion, omitted here for reasons of
  space, see\\ \url{http://bactra.org/weblog/232.html}.}

A rigorous application of the logic of error testing \cite{Mayo-error} would
now consider the probability of getting at least this good a fit to a
log-normal if the data were actually generated by a power law.  However, since
in this case the data were $e^{18481.51-17218.22} \approx 13$ million times
more likely under the log-normal distribution, any sane test would reject the
power-law hypothesis.

\subsection{Other Measures of Complexity}

Considerations of space preclude an adequate discussion of further complexity
measures.  It will have to suffice to point to some of the leading ones.  The
{\bold thermodynamic depth} of Lloyd and Pagels \cite{Lloyd-Pagels} measures
the amount of information required to specify a trajectory leading to a final
state, and is related both to departure from thermodynamic equilibrium and to
retrodiction \cite{TDCS}.  Huberman and Hogg \cite{Huberman-and-Hogg}, and
later Wolpert and Macready \cite{Wolpert-Macready-self-dissimilarity} proposed
to measure complexity as the {\em dissimilarity} between different levels of a
given system, on the grounds that self-similar structures are actually very
easy to describe.  (Say what one level looks like, and then add that all the
rest are the same!)  Wolpert and Macready's measure of self-dissimilarity is,
in turn, closely related to a complexity measure proposed by Sporns, Tononi and
Edelman \cite{Sporns-Tononi-Edelamn-connectivity-and-complexity,%
  Sporns-Tononi-Edelman-theoretical-neuroanatomy,Sporns-Tononi-in-complexity}
for biological networks, which is roughly the amount of information present in
higher-order interactions between nodes which is not accounted for by the
lower-order interactions.  Badii and Politi \cite{Badii-Politi} propose a
number of further {\bold hierarchical scaling complexities}, including one
which measures how slowly predictions converge as more information about the
past becomes available.  Other interesting approaches include the {\bold
  information fluctuation} measure of Bates and Shepard
\cite{Bates-Shepard-information-fluctuation}, and the predictability indices of
the ``school of Rome'' \cite{Predictability}.

\subsection{Relevance of Complexity Measures}

Why measure complexity at all? Suppose you are interested in the patterns of
gene expressions in tumor cells and how they differ from those of normal cells.
Why should you care if I analyze your data and declare that (say) healthy cells
have a more complex expression pattern?  Assuming you are not a numerologist,
the only reason you {\em should} care is if you can learn something from that
number --- if the complexity I report tells you something about the
thermodynamics of the system, how it responds to fluctuations, how easy it is
to control, etc.  A good complexity measure, in other words, is one which is
{\em relevant} to many other aspects of the system measured.  A bad complexity
measure lacks such relevance; a really bad complexity measure would be
positively misleading, lumping together things with no real connection or
similarity just because they get the same score.  My survey here has focused on
complexity measures which have some claim to relevance, deliberately avoiding
the large number of other measures which lack it \cite{DPF-JPC-why}.

\section{Guide to Further Reading}

\subsection{General}

There is no systematic or academically-detailed survey of the ``patterns'' of
complex systems, but there are several sound informal discussions: Axelrod and
Cohen \cite{Axelrod-Cohen-harnessing}, Flake \cite{Flake}, Holland
\cite{Holland-emergence} and Simon \cite{Simon-artificial}.  The book by Simon,
in particular, repays careful study.

On the ``topics'', the only books I can recommend are the ones by Boccara
\cite{Boccara-complex-systems} and Flake \cite{Flake}.  The former emphasizes
topics from physics, chemistry, population ecology and epidemiology, along with
analytical methods, especially from nonlinear dynamics.  Some sections will be
easier to understand if one is familiar with statistical mechanics at the level
of, e.g., \cite{Chandler-modern-stat-mech}, but this is not essential.  It does
not, however, describe any models of adaptation, learning, evolution, etc.
Many of those topics are covered in Flake's book, which however is written at a
much lower level of mathematical sophistication.

On foundational issues about complexity, the best available surveys
\cite{Badii-Politi,CMPPSS} both neglect the more biological aspects of the
area, and assume advanced knowledge of statistical mechanics on the part of
their readers.

\subsection{Data Mining and Statistical Learning}

There are now two excellent introductions to statistical learning and data
mining, \cite{Hand-Mannila-Smyth} and \cite{tEoSL}.  The former is more
interested in computational issues and the initial treatment of data; the
latter gives more emphasis to pure statistical aspects.  Both are recommended
unreservedly.  Baldi and Brunak \cite{Baldi-Brunak-bioinfo} introduces machine
learning via its applications to bioinformatics, and so may be especially
suitable for readers of this book.

For readers seriously interested in understanding the theoretical basis of
machine learning, \cite{Kearns-Vazirani} is a good starting point.  The work of
Vapnik \cite{Vapnik-nature,Vapnik-estimating,Vapnik-theory} is fundamental; the
presentation in his \cite{Vapnik-nature} is enlivened by many strong and
idiosyncratic opinions, pungently expressed.  \cite{Cristianini-Shawe-Taylor}
describes the very useful class of models called ``support vector machines'',
as well as giving an extremely clear exposition of key aspects of statistical
learning theory.  Those interested in going further will find that most of the
relevant literature is still in the form of journals --- {\em Machine
  Learning}, {\em Journal of Machine Learning Research} (free on-line at
\url{www.jmlr.org}), {\em Neural Computation} --- and especially annual
conference proceedings --- Computational Learning Theory (COLT), International
Conference on Machine Learning (ICML), Uncertainty in Artificial Intelligence
(UAI), Knowledge Discovery in Databases (KDD), Neural Information Processing
Systems (NIPS), and the regional versions of them (EuroCOLT, Pacific KDD,
etc.).

Much of what has been said about model selection could equally well have been
said about what engineers call {\bold system identification}, and in fact {\em
  is} said in good modern treatments of that area, of which
\cite{Nelles-system-identification} may be particularly recommended.

In many respects, data mining is an extension of exploratory data analysis; the
classic work by Tukey \cite{Tukey-EDA} is still worth reading.  No discussion
of drawing inferences from data would be complete without mentioning the
beautiful books by Tufte
\cite{Tufte-visual-display,Tufte-envisioning,Tufte-visual-explanations}.

\subsection{Time Series}

Perhaps the best all-around references for the nonlinear dynamics approach are
\cite{Kantz-Schreiber} and \cite{Abarbanel-analysis}.  The former, in
particular, succeeds in integrating standard principles of statistical
inference into the nonlinear dynamics method.  \cite{Sprott-on-time-series},
while less advanced than those two books, is a model of clarity, and contains
an integrated primer on chaotic dynamics besides.  Ruelle's little book
\cite{Ruelle-Lincei} is {\em much} more subtle than it looks, full of deep
insights.  The SFI proceedings volumes
\cite{Casdagli-Eubank,time-series-prediction} are very worthwhile.  The
journals {\em Physica D}, {\em Physical Review E} and {\em Chaos} often have
new developments.

From the statistical wing, one of the best recent textbooks is
\cite{Shumway-Stoffer}; there are many, many others.  That by Durbin and
Koopman \cite{Durbin-Koopman-state-space-methods} is particularly strong on the
state-space point of view.  The one by \cite{Azencott-Dacunha-Castelle}
Azencott and Dacunha-Castelle is admirably clear on both the aims of time
series analysis, and the statistical theory underlying classical methods;
unfortunately it typography is less easy to read than it should be.
\cite{Taniguchi-Kakizawa} provides a comprehensive and up-to-date view of the
statistical theory for modern models, including strongly non-linear and
non-Gaussian models.  While many of the results are directly useful in
application, the proofs rely on advanced theoretical statistics, in particular
the geometric approach pioneered by the Japanese school of Amari {\em et al.}
\cite{Amari-Nagaoka}, This {\bold information geometry} has itself been applied
by Ay to the study of complex systems
\cite{Ay-pragmatic-structuring,Ay-information-geometry-on-complexity}.

At the interface between the statistical and the dynamical points of view,
there is an interesting conference proceedings
\cite{Cutler-Kaplan-nonlinear-dynamics-and-time-series} and a useful book by
Tong \cite{Tong-nonlinear-time-series}.  Pearson's book
\cite{Pearson-discrete-time-dynamic} on discrete-time models is very good on
many important issues related to model selection, and exemplifies the habit of
control theorists of cheerful stealing whatever seems helpful.

\paragraph{Filtering}

Linear filters are well-described by many textbooks in control theory
(e.g. \cite{Stengel-optimal-control}), signal processing, time series analysis
(e.g. \cite{Shumway-Stoffer}) and stochastic dynamics
(e.g. \cite{Honerkamp-stochastic}).

\cite{Eyink-variational-optimal-estimation} provides a readable introduction to
optimal nonlinear estimation, draws interesting analogies to non-equilibrium
statistical mechanics and turbulence, {\em and} describes a reasonable
approximation scheme.  \cite{Ahmed-filtering} is an up-to-date textbook,
covering both linear and nonlinear methods, and including a concise exposition
of the essential parts of stochastic calculus.  The website run by
R. W. R. Darling, \url{www.nonlinearfiltering.webhop.net}, provides a good
overview and extensive pointers to the literature.

\paragraph{Symbolic Dynamics and Hidden Markov Models}

On symbolic dynamics, formal languages and hidden Markov models generally, see
\cite{Badii-Politi}.  \cite{Lewis-Papadimitriou-computation} is a good first
course on formal languages and automata theory.  Charniak is a very readable
introduction to grammatical inference. \cite{Lind-Marcus} is an advanced
treatment of symbolic dynamics emphasizing applications; by contrast,
\cite{Kitchens} focuses on algebraic, pure-mathematical aspects of the subject.
\cite{Beck-Schlogl} is good on the stochastic properties of symbolic-dynamical
representations.  Gershenfeld \cite{Gershenfeld-modeling} gives a good
motivating discussion of hidden Markov models, as does Baldi and Brunak
\cite{Baldi-Brunak-bioinfo}, while \cite{Elliott-et-al-HMM} describes advanced
methods related to statistical signal processing.  Open-source code for
reconstructing causal-state models from state is available from
\url{http://bactra.org/CSSR}.

\subsection{Cellular Automata}

\paragraph{General}

There is unfortunately no completely satisfactory unified treatment of cellular
automata above the recreational.  Ilachinski \cite{Ilachinski-discrete}
attempts a general survey aimed at readers in the physical sciences, and is
fairly satisfactory on purely mathematical aspects, but is more out of date
than its year of publication suggests.  Chopard and Droz
\cite{Chopard-Droz-text} has good material on models of pattern formation
missing from Ilachinski, but the English is often choppy.  Toffoli and Margolus
\cite{Toffoli-Margolus} is inspiring and sound, though cast on a piece of
hardware and a programming environment which are sadly no longer supported.
Much useful material on CA modeling has appeared in conference proceedings
\cite{Farmer-Toffoli-Wolfram-CA-conference,Gutowitz-CA-conference,%
  Cellular-Automata-and-Modeling}.

\paragraph{CA as Self-Reproducing Machines}

The evolution of CA begins in \cite{von-Neumann-self-reproducing}, continues
in \cite{Burks-essays}, and is brought up to the modern era in
\cite{Poundstone-recursive}; the last is a beautiful, thought-provoking and
modest book, sadly out of print.  The modern era itself opens with
\cite{ALife-1}.

\paragraph{Mathematical and Automata-Theoretic Aspects}

Many of the papers in \cite{Wolfram-CA-and-complexity} are interesting.
Ilachinski \cite{Ilachinski-discrete}, as mentioned, provides a good survey.
The Gutowitz volume \cite{Gutowitz-CA-conference} has good material on this
topic, too.  \cite{new-constructions-in-CA} is up-to-date.

\paragraph{Lattice gases}

\cite{Rothman-Zaleski-text} is a good introduction,
\cite{Rivet-Boon-lattice-gas-hydro} somewhat more advanced.  The pair of
proceedings edited by Doolen
\cite{Doolen-lattice-gas-for-pdes,Doolen-lattice-gas-methods} describe many
interesting applications, and contain useful survey and pedagogical articles.
(There is little overlap between the two volumes.)

\subsection{Agent-Based Modeling}

There do not seem to be any useful textbooks or monographs on agent-based
modeling.  The {\em Artificial Life} conference proceedings, starting with
\cite{ALife-1}, were a prime source of inspiration for agent-based modeling,
along with the work of Axelrod \cite{Axelrod-evol-of-coop}.
\cite{Varela-Bourgine-practice-of-autonomous} is also worth reading.  The
journal {\em Artificial Life} continues to be relevant, along with the {\em
  From Animals to Animats} conference series.  Epstein and Axtell's book
\cite{Epstein-Axtell} is in many ways the flagship of the ``minimalist''
approach to ABMs; while the arguments in its favor (e.g.,
\cite{Epstein-generative-social-science,Macy-Willer-factors-to-actors}) are
often framed in terms of social science, many apply with equal force to
biology\footnote{In reading this literature, it may be helpful to bear in mind
  that by ``methodological individualism'', social scientists mean roughly what
  biologists do by ``reductionism''.}.  \cite{Agents-and-GIS} illustrates how
ABMs can be combined with extensive real-world data.  Other notable
publications on agent-based models include \cite{Kohler-Gumerman-dynamics},
spanning social science and evolutionary biology;
\cite{Bonabeau-on-morphogenesis} on agent-based models of morphogenesis; and
\cite{Camazine-et-al-self-org-in-bio} on biological self-organization.

\cite{Budd-understanding-oop} introduces object-oriented programming and the
popular Java programming language at the same time; it also discusses the roots
of object-orientation in computer simulation.  There are many, many other books
on object-oriented programming.

\subsection{Evaluating Models of Complex Systems}

Honerkamp \cite{Honerkamp-stochastic} is great, but curiously almost unknown.
Gershenfeld \cite{Gershenfeld-modeling} is an extraordinary readable
encyclopedia of applied mathematics, especially methods which can be used on
real data.  Gardiner \cite{Gardiner-handbook} is also useful.

\paragraph{Monte Carlo}

The old book by Hammersley and Handscomb \cite{Hammersley-Handscomb} is
concise, clear, and has no particular prerequisites beyond a working knowledge
of calculus and probability.  \cite{MEJN-on-Monte-Carlo} and
\cite{MacKeown-simulation} are both good introductions for readers with some
grasp of statistical mechanics.  There are also very nice discussions in
\cite{Honerkamp-stochastic,tEoSL,Bremaud-markov-chains}.  Beckerman
\cite{Beckerman-adaptive} makes Monte Carlo methods the starting point for a
fascinating and highly unconventional exploration of statistical mechanics,
Markov random fields, synchronization and cooperative computation in neural and
perceptual systems.

\paragraph{Experimental design}
Bypass the cookbook texts on standard designs, and consult Atkinson and Donev
\cite{Atkinson-Donev} directly.

\paragraph{Ecological inference}
\cite{King-ecological-inference} is at once a good introduction, and the source
of important and practical new methods.

\subsection{Information Theory}

Information theory appeared in essentially its modern form with Shannon's
classic paper \cite{Shannon-1948}, though there had been predecessors in both
communications \cite{Hartley-1928} and statistics, notably Fisher (see Kullback
\cite{Kullback-info-theory-and-stats} for an exposition of these notions), and
similar ideas were developed by Wiener and von Neumann, more or less
independently of Shannon \cite{Wiener-cybernetics}.  Cover and Thomas
\cite{Cover-and-Thomas} is, deservedly, the standard modern textbook and
reference; it is highly suitable as an introduction, and handles almost every
question most users will, in practice, want to ask.  \cite{Gray-entropy} is a
more mathematically rigorous treatment, now free on-line.  On neural
information theory, \cite{Spikes-book} is seminal, well-written, still very
valuable, and largely self-contained.  On the relationship between physics and
information, the best reference is still the volume edited by Zurek
\cite{complexity-entropy-physics-of-info}, and the thought-provoking paper by
Margolus.

\subsection{Complexity Measures}

The best available survey of complexity measures is that by Badii and Politi
\cite{Badii-Politi}; the volume edited by Peliti and Vulpiani
\cite{Peliti-Vulpiani}, while dated, is still valuable.  Edmonds
\cite{Edmonds-complexity-bibliography} is an online bibliography, fairly
comprehensive through 1997.  \cite{CMPPSS} has an extensive literature review.

On Kolmogorov complexity, see Li and Vitanyi \cite{Li-and-Vitanyi-1993}.  While
the idea of measuring complexity by the length of descriptions is usually
credited to the trio of Kolmogorov, Solomonoff and Chaitin, it is implicit in
von Neumann's 1949 lectures on the ``Theory and Organization of Complicated
Automata'' \cite[Part I, especially pp. 42--56]{von-Neumann-self-reproducing}.

On MDL, see Rissanen's book \cite{Rissanen-SCiSI}, and Gr\"unwald's lecture
notes \cite{Grunwald-tutorial-on-MDL}.  Vapnik \cite{Vapnik-nature} argues that
when MDL converges on the optimal model, SRM will too, but he assumes
independent data.

On statistical complexity and causal states, see \cite{CMPPSS} for a
self-contained treatment, and \cite{CRS-thesis} for extensions of the theory.

\begin{acknowledgments}

This work has been supported by a grant from the James S. McDonnell Foundation,
and was finished while enjoying the hospitality of the Institut des
Syst{\`e}mes Complexes, the Laboratoire de l'Informatique du Parall{\'e}isme
and the Exystence Thematic Institute on Discrete and Computational Aspects of
Complex Systems at the {\'E}cole Normale Superiere de Lyon.  I am grateful to
Michel Morvan and Cris Moore, organizers of the ``Science et Gastronomie 2003''
workshop, for allowing me to present some of this material there, and to my
fellow participants for their critical comments.  It is a pleasure to
acknowledge discussions with Dave Albers, Satinder Singh Baveja, Philippe
Binder, Sven Bruckener, Sandra Chapman, Markus Christen, Michael Cohen, Jim
Crutchfield, Gunther Eble, Dave Feldman, Walter Fontana, Peter Grassberger, Rob
Haslinger, John Holland, Herbert Jaeger, J\"urgen Jost, Michael Lachmann, Alex
Lancaster, Norm Margolus, Cris Moore, Mark Newman, Scott Page, Mitchell Porter,
Kristina Shalizi, Carl Simon, Eric Smith, Ricard Sol{\'e}, Bill Tozier, Erik
van Nimwegen and Nick Watkins.  Special thanks to Prof.\ Cohen for permission
to take a triangle he drew and add a corner to it; to Profs.\ Drenzer and
Farrell for permission to use their data on weblogs; to Bill Tozier for
suggesting that the leitmotifs of complex systems are analysis patterns, and
for advice on agents; to Kara Kedi for daily validation of von Neumann's
remarks on the complexity of cats; to Mathew Dafilis and Nigel Phillips for
spotting misprints; to Kristina Shalizi for assistance on linear models of time
series and for careful reading of a draft; and to Kristina and Rob for
resisting the idea that there {\em is} such a thing as complex systems science.
Finally, I wish to acknowledge the patience of the editors.

\end{acknowledgments}

\bibliographystyle{unsrt}
\chapbblname{methods}
\chapbibliography{locusts}

\end{document}